\RequirePackage{fix-cm}
\documentclass[twocolumn,epjc3]{svjour3}  
\smartqed  

\RequirePackage{graphicx}
\RequirePackage{dcolumn}
\RequirePackage{bm}
\RequirePackage{color} 
\RequirePackage{xcolor} 
\RequirePackage{color} 
\RequirePackage{graphics}
\RequirePackage{float}
\RequirePackage{bm}        
\RequirePackage{amssymb}   
\RequirePackage{slashed}   
\RequirePackage{amsmath}   
\RequirePackage{verbatim}  
\RequirePackage{subfig}
\RequirePackage{mathptmx}      
\RequirePackage{multirow}
\RequirePackage{flushend}
\RequirePackage[numbers,sort&compress]{natbib}
\RequirePackage[colorlinks,citecolor=blue,urlcolor=blue,linkcolor=blue]{hyperref}

\begin{document}


\title{Study of the effects of external imaginary electric field and chiral chemical potential on quark matter}




\author{Ji-Chong Yang\thanksref{e1,addr1,addr2}
        \and
        Xin Zhang\thanksref{e2,addr1,addr2}
        \and
        Jian-Xing Chen\thanksref{cr,e3,addr1,addr2}
}

\institute{Department of Physics, Liaoning Normal University, Dalian 116029, China\label{addr1} 
           \and
          Center for Theoretical and Experimental High Energy Physics, Liaoning Normal University, Dalian 116029, China\label{addr2}}

\thankstext[$\star$]{cr}{Corresponding author}
\thankstext{e1}{e-mail: yangjichong@lnnu.edu.cn}
\thankstext{e2}{e-mail: zx1280475084@163.com}
\thankstext{e3}{e-mail: 13614090213@163.com}

\date{Received: date / Revised version: date}

\maketitle

\begin{abstract}
The behavior of quark matter with both external electric field and chiral chemical potential is theoretically and experimentally interesting to consider. 
In this paper, the case of simultaneous presence of imaginary electric field and chiral chemical potential is investigated using the lattice QCD approach with $N_f=1+1$ dynamical staggered fermions. 
We find that overall both the imaginary electric field and the chiral chemical potential can exacerbate chiral symmetry breaking, which is consistent with theoretical predictions. 
However we also find a non-monotonic behavior of chiral condensation at specific electric field strengths and chiral chemical potentials. 
The behavior of Polyakov loop in the complex plane is not significantly affected by chiral chemical potential in the region of the parameters considered.
\end{abstract}


\maketitle


\section{\label{sec:level1}Introduction}

In the quark-gluon-plasma phase chirality imbalance of QCD is expected~\cite{Adler:1969gk,Bell:1969ts,Andrianov:2016qgy}, so it is of interest to study how chirality imbalance modifies the structure of QCD.
A convenient way to treat quark matter with net chirality is to introduce an chiral~(axial) chemical potential, $\mu_5$, conjugated to chiral density, the difference between the number of right- and left-handed fermions, $n_5 = n_R-n_L$. 
The chiral chemical potential can be introduced to study the topological changing transitions~\cite{Klinkhamer:1984di,Kuzmin:1985mm,McLerran:1990de,Moore:1999fs,Shuryak:2002qz}, where the effect of topological charges can be mimicked by an effective $\theta$ angle, and $\mu _5$ can be interpreted as the time derivative of the $\theta$ angle as $\mu _5 = \theta'(t)/(2N_f)$, where $t$ is the time coordinate, and $N_f$ is the number of flavors~\cite{Fukushima:2008xe,Chernodub:2011fr}. 
One of its most important applications is the chiral magnetic effect~(CME)~\cite{Kharzeev:2004ey,Fukushima:2008xe}.

Various effective models such as the Nambu-Jona-Lasinio~(NJL) model~\cite{Yu:2015hym}, the Polyakov loop extended NJL model~\cite{Fukushima:2010fe,Gatto:2011wc}, the linear sigma model coupled to the Polyakov loop~\cite{Chernodub:2011fr,Ruggieri:2011xc} and chiral perturbation theory~\cite{Espriu:2020dge} are used to investigate effects of chiral chemical potential, including the effects of chiral chemical potential on chiral magnetic effect, chiral phase transition, critical end point in quark matter, etc. 
The chirality imbalanced hot and dense strongly interacting matter is also studied by means of the Dyson-Schwinger equations~\cite{Shi:2020uyb}. 
Catalysis effect of dynamical chiral symmetry breaking by chiral chemical potential is observed and chiral charge density generally increases with temperature, quark number chemical potential and chiral chemical potential. 

The effect of an external electric field on the QCD vacuum structure is also a very interesting subject because of academic and realistic reasons. 
In the relativistic heavy-ion collisions, the electric fields can be generated owing to the event-by-event fluctuations or in asymmetric collisions like $Cu+Au$ collision, and the strength of the electric fields can be roughly of the same order as the magnetic fields~\cite{Bzdak:2011yy,Deng:2012pc,Bloczynski:2012en,Hirono:2012rt,Deng:2014uja,Voronyuk:2014rna}.
It has been shown that, the electric field restores the chiral symmetry~\cite{Babansky:1997zh,Klevansky:1989vi,Suganuma:1990nn,Tavares:2019mvq,Cao:2015dya,Ruggieri:2016xww,Ruggieri:2016jrt,Ruggieri:2016lrn}, and $n_5$ can be affected by the external electric field and magnetic field. 
The impact of electric and magnetic fields on chirality has also been studied using lattice approach~\cite{Brandt:2022jfk}.

In lattice QCD approach, the chiral chemical potential was introduced to study the CME~\cite{Yamamoto:2011gk}, and has been studied in various previous works~\cite{Yamamoto:2011ks,Braguta:2014ira,Braguta:2015zta,Kotov:2015hxr,Braguta:2019pxt,Astrakhantsev:2019wnp}.
It is confirmed that $n_5$ is proportional to the magnetic field and to the chiral chemical potential.
In the case that only $\mu _5$ is presented, both string tension and chiral susceptibility grow with the chiral chemical potential.
In this paper, we consider the case where both the electric and chiral chemical potential are presented.

The remainder of this paper is organized as follows, in Sec.~\ref{sec:2} the action on the lattice is briefly introduced, the numerical results are shown in Sec.~\ref{sec:3}, the Sec.~\ref{sec:4} is a summary.

\section{\label{sec:2}The model}

In this paper, we consider the case of an external uniform electric field at ${\bf z}$ direction and with the chiral chemical potential. 
The electric gauge field can be written as $A^{\rm EM}_{\mu}=(-E_z z, 0, 0, 0)$ in the axial gauge, the superscript `EM' is added to distinguish with the QCD gauge field.
The Lagrangian with one massless fermion is
\begin{equation}
\begin{split}
&\mathcal{L}_{q}=\bar{\psi} _q\slashed{\partial} _{\mu}\psi _q+ \bar{\psi} _q i \slashed{A}\psi _q -i Q_q e E_z z \bar{\psi}_q \gamma _0 \psi _q - \mu_5 \bar{\psi}_q \gamma _0\gamma _5 \psi,
\end{split}
\label{eq.2.1}
\end{equation}
where the last term corresponds to the chiral chemical potential.
The fermion action is,
\begin{equation}
\begin{split}
&S_q=\int d^4x^E \left(\bar{\psi} \sum _{j=1}^4 \gamma _j^E \partial _j \psi + \sum _{j=1}^4 \bar{\psi} i g \gamma _j^E A_j\psi \right.\\
&\left.-i Q_q e E_z z \bar{\psi} \gamma _4^E \psi- \mu_5 \bar{\psi}_q \gamma _4^E \gamma _5 \psi\right),
\end{split}
\label{eq.2.2}
\end{equation}
where $\gamma _{j=1,2,3}=i \gamma _j^E$, and $\gamma _4^E = \gamma _0$.

The chiral chemical potential tends to break the chiral symmetry.
Meanwhile, the electric field tends to break chiral condensation whether the chiral imbalance is taken into account or not.
Apart from that, chirality imbalance can be affected when homogeneous parallel electric field and magnetic field are presented~\cite{Ruggieri:2016lrn}.
Therefore, the case of quark matter with chiral chemical potential and external electric field is interesting.

It is known that, using lattice approach, there is notorious `sign problem' in the case of external real electric field, except for the case that $u$ and $d$ quarks have opposite charges~\cite{Yamamoto:2012bd}.
To void the `sign problem', analytical continuation is often used to study the case of electric field, which is found to be reliable~\cite{Shintani:2006xr,Alexandru:2008sj,DElia:2012ifm,Fiebig:1988en,Christensen:2004ca,Engelhardt:2007ub,Endrodi:2022wym,Endrodi:2021qxz,Yang:2022zob}.

Except for the boundaries, in axial gauge the presence of the external electric field can be viewed as a stacking of volumes with different imaginary chemical potentials $\mu =Q_q e E_z z$ extending the ${\bf z}$-axis.
The case of a homogeneous imaginary chemical potential and the corresponding Roberge-Weiss~(R-W) transition~\cite{Roberge:1986mm} has been studied~\cite{Philipsen:2014rpa,Wu:2013bfa,Wu:2014lsa,Cuteri:2015ayx,Bonati:2018fvg,Cuteri:2022vwk}.
Similar R-W transition is also found in the case of electric field~\cite{Yang:2022zob}~(similar behavior is also found in the case of magnetic field~\cite{DElia:2016kpz}).
In this paper, the behavior of Polyakov loop at different $z$ coordinate in the complex plane is also studied.

On the lattice, the discretized action with staggered fermions, and without the electric field and the chiral chemical potential can be written as~\cite{Kogut:1974ag,Wilson:1974sk,Gattringer:2010zz},
\begin{equation}
\begin{split}
&S_G=\frac{\beta}{N_c}\sum _n \sum _{\mu>\nu}{\rm Retr}\left[1-U_{\mu\nu}(n)\right],\\
&S_q=\sum _n \left(\sum _{\mu}\sum _{\delta = \pm \mu}\bar{\chi}(n) U_{\delta}(n)\eta _{\delta}(n)\chi (n+\delta) +2am\bar{\chi}\chi \right),
\end{split}
\label{eq.2.3}
\end{equation}
where $a$ is the lattice spacing, $\beta=2N_c/g^2_{\rm YM}$ with $g_{\rm YM}$ the coupling strength of the gauge fields to the quarks, $m$ is the fermion mass, $U_{\mu}=e^{iaA_{\mu}}$, $\eta _{\mu}(n)=(-1)^{\sum _{\nu<\mu}n_{\nu}}$ and $U_{-\mu}(n)=U_{\mu}^{\dagger}(n-\mu)$, $\eta _{-\mu}=-\eta _{\mu}(n-\mu)$.
We will introduce the electric field and the chiral chemical potential in the following subsections.

\subsection{\label{sec:2.1}Electric field}

To avoid the `sign problem', in this paper a Wick rotation is performed to the electric field, in other words, a substitution $A^{\rm EM}_0 \to iA^{\rm EM}_4$ is applied.
The result of the substitution corresponds to an imaginary electric field, which is also known as an `Euclidean' or `classical' electric field.

The electric field is introduced as a $U(1)$ phase in this paper, i.e., a $U(1)$ phase corresponds to the external electric field is added to the gauge links connecting bilinear terms of $\chi$, the fermion action without chiral chemical potential is then,
\begin{equation}
\begin{split}
&S^{\rm EM}_q=\\
&\sum _n \left(\sum _{\mu}\sum _{\delta = \pm \mu}\bar{\chi}(n) U_{\delta}(n)V_{\delta}(n)\eta _{\delta}(n)\chi (n+\delta) +2am\bar{\chi}\chi \right),
\end{split}
\label{eq.2.4}
\end{equation}
where $V_{\mu}(n)=e^{iaQ_qA^{\rm EM}_{\mu}(n)}$, $V_{-\mu}(n)=V^*_{\mu}(n-\mu)$ which is similar as the definition of $U_{\mu}(n)$.

To ensure gauge invariance of the external electric field and the translational invariance at the same time, a twisted boundary condition is applied~\cite{DElia:2012ifm,Damgaard:1988hh,Al-Hashimi:2008quu,Buividovich:2009bh}, in this paper we use
\begin{equation}
\begin{split}
&f=Q_qa^2F=\frac{2k\pi}{L_{\mu}L_{\nu}},\;\;\;k\in \mathbb{Z},\\
&V_{\nu}=e^{i f n_{\mu}},\;\;V_{\mu}(n_{\mu}=L_{\mu})=e^{-ifL_{\mu}n_{\nu}}.\\
\end{split}
\label{eq.2.5}
\end{equation}
where $L_{\mu}$ is the extent at direction $\mu$.

In this paper, the origin of the axis is set to be the middle of the spatial volume and at $n_{\tau}=1$.
Therefore, $V_{\mu}(n)=1$ except for
\begin{equation}
\begin{split}
&V_{\tau}(n)=e^{-i a Q_q e E_z z},\;\;V_z(a^{-1}z=\frac{L_z}{2}-1)=e^{i a Q_q e E_z L_z \tau},\\
\end{split}
\label{eq.2.6}
\end{equation}

The strength of electric field is quantized and is decided by $L_z\times L_{\tau}$.
In this paper, the extents of the lattice is $L_x\times L_y\times L_z\times L_{\tau}=8\times 8\times 24\times 6$, with $|Q_q|=1/3$, it is required that $a^2 e\Delta E_z=2\pi/(|Q_q|L_zL_{\tau})=\pi / 24$.
Besides, it is also required that $\exp(-iQ_qa^2eE_z)$ to be an approximation of $1-iQ_qa^2eE_z$, as a consequence, for $a^2eE_z\sim \mathcal{O}(1)$ or larger, the result suffer from strong discretization errors.
To keep our result qualitatively reliable, we use $a^2 e E_z = k\times \pi / 24$ where $k$ is an integer satisfying $0\leq k \leq 8$.

\subsection{\label{sec:2.2}Chiral chemical potential}

By using the definition of the staggered quarks, $\Psi^{\alpha a}(h)=\frac{1}{8}\sum _{\mu}\sum _{\delta _{\mu}=0,1}\Gamma _{\delta}^{\alpha a}\psi(h+\delta)$ where $\Gamma _{\delta} = (\gamma _1)^{\delta _1}(\gamma _2)^{\delta _2}(\gamma _3)^{\delta _3}(\gamma _4)^{\delta _4}$~\cite{Kluberg-Stern:1983lmr,Morel:1984di}, and $\chi = a^2 \psi /\sqrt{2a}$, it can be shown that~\cite{Yang:2023vsw}
\begin{equation}
\begin{split}
&(2a)^4\sum _h \bar{\Psi}(h)\gamma _5\gamma _4\otimes I \Psi(h) + \mathcal{O}(a) =a^4S_q^{\mu_5},\\
&S_q^{\mu_5}= \frac{1}{8}\sum _n \sum _{\vec{s}_{x,y,z}=\pm 1}\left(-\eta _1(n)\eta _2(n)\eta _3(n)\right.\\
&\left.\times \bar{\chi}(n)U(n,n+\vec{s})V(n,n+\vec{s})\chi(n+\vec{s})\right)\\
\end{split}
\label{eq.2.7}
\end{equation} 
where $\vec{s}$ is an offset vector pointing to diagonals of elementary cubes in the ${\bf x}-{\bf y}-{\bf z}$ space.
The sum over $h$ in the left hand side of Eq.~(\ref{eq.2.7}) is a sum over even sites~($2^4$ hypercubes).
The $U,V(n,n+\vec{s})$ are Wilson lines of $SU(3)$ and $U(1)$ gauge fields connecting $n$ and $n+\vec{s}$.
For a three dimensional cube, there are six shortest paths connecting $n$ and $n+\vec{s}$, similar as in Refs.~\cite{Braguta:2014ira,Braguta:2015zta}, we use the average of the Wilson lines along the six paths.

In this paper, the $\mu_5$ term is introduced in a linear form in the action, instead of exponential factors on the links.
In this case, there will be $\mu_5$-dependent additive divergences in the observables.
The case of exponential form is treated using Taylor-expansion and is discussed in the study of the anomalous transport phenomena~\cite{Velasco:2022gaw}.

The fermion action with both the electric field and chiral chemical potential considered is then $S_q=S_q^{\rm EM}+S_q^{\mu _5}$.

\section{\label{sec:3}Numerical results}

\begin{figure}[htbp]
\begin{center}
\includegraphics[width=0.8\hsize]{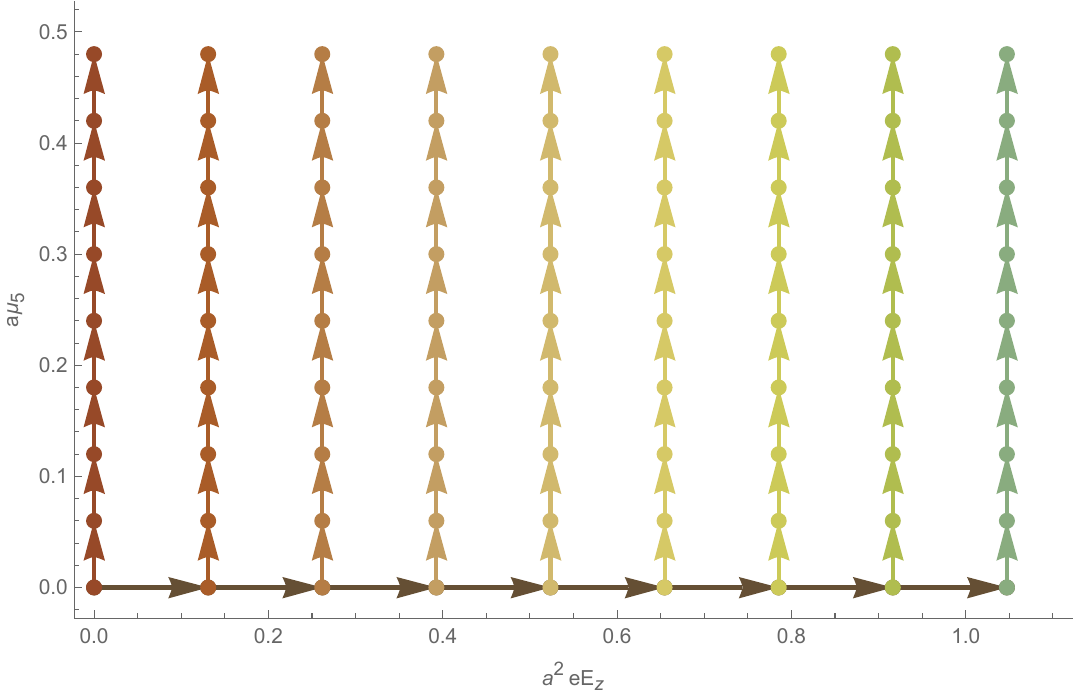}\\
\caption{\label{fig:simulationpath}
The parameters in the simulation.
In the case of $a^2eE_z=a\mu_5$, $3200$ trajectories are simulated and the last $2900$ configurations are measured, for the other cases, $3000$ trajectories and the last $2900$ configurations are measured.
At first, $200+3000\times 9$ trajectories are simulated with sequentially growing $a^2eE_z=k\pi/24$ for $k=0,1,2,\ldots, 8$, which corresponds to the horizontal line.
Then, starting with the configuration at $a^2eE_z=k\pi/24$, $3000\times 8$ trajectories are simulated with sequentially growing $a\mu_5=0.06n$ for $n=1,2,3,\ldots, 8$, which correspond to the vertical lines.
}
\end{center}
\end{figure}

In this paper, we simulate on a lattice with $N_x\times N_y \times N_z\times N_t=8\times 8 \times 24 \times 6$, and with $N_f=1+1$ staggered fermions with $am = 0.1$.
The dynamic fermions with same masses are charged as $Q_u =2/3$ and $Q_d=-1/3$.
For one flavour, the `forth root trick' is applied to remove the taste degeneracies, and the rational hybrid Monte Carlo~\cite{Clark:2006wp} is used.
For each $\beta$, at first, $200+3000\times 9$ trajectories are simulated.
The first $200$ trajectories are discarded for thermalization, then $3000\times 9$ trajectories are simulated with sequentially growing $a^2eE_z=k\pi/24$ for $k=0,1,2,\ldots, 8$.
Then, we simulate starting with the configuration at $a^2eE_z=k\pi/24$, and $3000\times 8$ trajectories are simulated with sequentially growing $a\mu_5=0.06n$ for $n=1,2,3,\ldots, 8$.
That is, for each $a^2eE_z$ and $a\mu_5$, $3000$ configurations are obtained, where the first $100$ trajectories are discarded for thermalization and $2900$ configurations are measured.
A summarize of the change of parameters in the simulation is shown in Fig.~\ref{fig:simulationpath}.

In this paper, the statistical error is calculated as $\sigma=\sigma _{\rm jk} \sqrt{2\tau _{\rm ind}}$~\cite{Gattringer:2010zz}, where $\sigma _{\rm jk}$ is statistical error calculated using `jackknife' method, and $2\tau _{\rm ind}$ is the separation of molecular dynamics time units~(TU) such that the two configurations can be regarded as independent.
$\tau_{\rm ind}$ is calculated by using `autocorrelation' with $S=1.5$~\cite{Wolff:2003sm} on the bare chiral condensation of $u$ quark.
A brief introduction of the calculation of $\tau _{int}$ is established in \ref{sec:ap0}.

\subsection{\label{sec:3.1}Matching}

In this paper, we focus on the cases of $\beta=5.3$ and $\beta = 5.4$.
In Ref.~\cite{Yang:2022zob}, these parameters are studied on a $12^3 \times 48$ lattice, at $\beta = 5.3$ and $\beta = 5.4$, $a^{-1}=1215\pm 12 \;{\rm MeV}$ and $a^{-1}=1623\pm 16 \;{\rm MeV}$, respectively.
Therefore, $\beta = 5.3$ and $\beta = 5.4$ correspond to $T=202.5\pm 2.0 \;{\rm MeV}$ and $T=270.5\pm 2.7\;{\rm MeV}$, respectively.

$am=0.1$ is a relatively heavy quark mass.
Heavier quark masses require relatively fewer computational resources, and are therefore this setup is only used for studies with qualitative conclusions.
In the case of $12^3\times 48$ lattice, the masses of mesons are estimated using the method in Ref.~\cite{Gottlieb:1988gr}, where we use the Los Alamos gauge fixing method~\cite{Cucchieri:1996jm,Paciello:1991bd}.
The masses of $\pi$ and $\rho$ mesons correspond to the forward direction propagators in the pseudoscalar and vector tensor channels.
The effective masses of $\pi$ and $\rho$ mesons are $0.967\pm 0.003$ and $1.484\pm 0.008$ at $\beta=5.3$, and $0.985\pm 0.003$ and $1.452\pm 0.008$ at $\beta=5.4$, respectively, which are close to the effective mass when $am=0.1$ in Ref.~\cite{Gottlieb:1988gr}.
This indicates that, at $\beta=5.3$, the masses of $\pi$ and $\rho$ mesons are $1175\pm 12\;{\rm MeV}$ and $1803\pm 20\;{\rm MeV}$, at $\beta=5.4$, the masses of $\pi$ and $\rho$ mesons are $1599\pm 16\;{\rm MeV}$ and $2357\pm 27\;{\rm MeV}$, respectively.

The breaking/restoration of chiral symmetry phase transition is also studied on a $12^3\times 6$ lattice~\cite{Yang:2022zob}.
The pseudo critical temperature is $T=229\;{\rm MeV}$ which corresponds to $\beta = 5.34$.
Therefore, in this paper, the case of $\beta = 5.3$ corresponds to chiral symmetry breaking phase, and $\beta = 5.4$ corresponds to chiral symmetry restoring phase, and both cases are close to the pseudo critical temperature.

In Ref.~\cite{Yang:2022zob}, a transition related to the R-W phase is observed. 
There is a phases transition at high temperature and with large electric field, where the winding number of the function of Polyakov loop along $z$ coordinate  at the complex plane changes from none-zero to zero, and this is accompanied by the phenomena that the chiral condensation start to oscillate along the ${\bf z}$-axis, and the local charge density becomes non-zero.
However, this transition does not occur at $\beta=5.3\;(T=202.5\;{\rm MeV})$, nor does it occur at $\beta=5.4\;(T=270.5\;{\rm MeV})$ and $a^2eE_z\leq \pi / 3$.
Therefore, for the parameters used in this paper, this transition is irrelevant.

\subsection{\label{sec:3.2}Chiral condensation}

The chiral condensation is related to the chiral symmetry.
It has been shown in Ref.~\cite{Yang:2022zob} that, the imaginary electric field will increase the chiral condensation.
Meanwhile, the chiral chemical potential will also increase the chiral condensation~\cite{Braguta:2014ira,Braguta:2015zta,Braguta:2019pxt}.
The behavior of the chiral condensation with both the electric field and chiral chemical potential is of interest.
Besides, how the chiral charge density $\bar{n}_5=n_5/V$ is affected by the presences of electric field and chiral chemical potential is also investigated, where $\bar{n}_5=\langle \bar{\Psi}\gamma _5\gamma _4\Psi\rangle/V$ with $V$ denotes the volume.

The measured quantities correspond to $\langle \bar{\Psi} \Psi \rangle$ and $\bar{n}_5$ are
\begin{equation}
\begin{split}
&c_q=\frac{1}{4}\times \frac{1}{V}\sum _n\langle \bar{\chi}_q(n)\chi_q(n)\rangle,\\
&\bar{n}_{5,q}=\frac{1}{4}\times \frac{1}{V}\times \frac{1}{8}\sum _n \sum _{\vec{s}_{x,y,z}=\pm 1}\langle \bar{\chi}_q(n) \left(-\eta _1(n)\eta _2(n)\eta _3(n)\right)\\
&\times U(n,n+\vec{s})V(n,n+\vec{s})\chi_q(n+\vec{s})\rangle,\\
\end{split}
\label{eq.3.1}
\end{equation}
where four in the denominator is the number of taste $N_{taste}=4$, $V=L_x\times L_y\times L_z\times L_{\tau}$.

In Ref.~\cite{Yang:2022zob}, it has been found that at high temperatures the chiral condensation has an oscillation over the ${\bf z}$ direction.
Meanwhile, at high temperatures, there is also imaginary charge condensation which depends on ${\bf z}$ coordinate induced by imaginary electric field~\cite{Endrodi:2022wym,Endrodi:2021qxz,Yang:2022zob}.
However, the temperature and strength of the electric field used in this paper is not high enough to see the oscillation and charge condensation, therefore we do not consider the ${\bf z}$ dependence of the condensations, and the charge condensation is not considered.

\begin{figure}[htbp]
\begin{center}
\includegraphics[width=0.48\hsize]{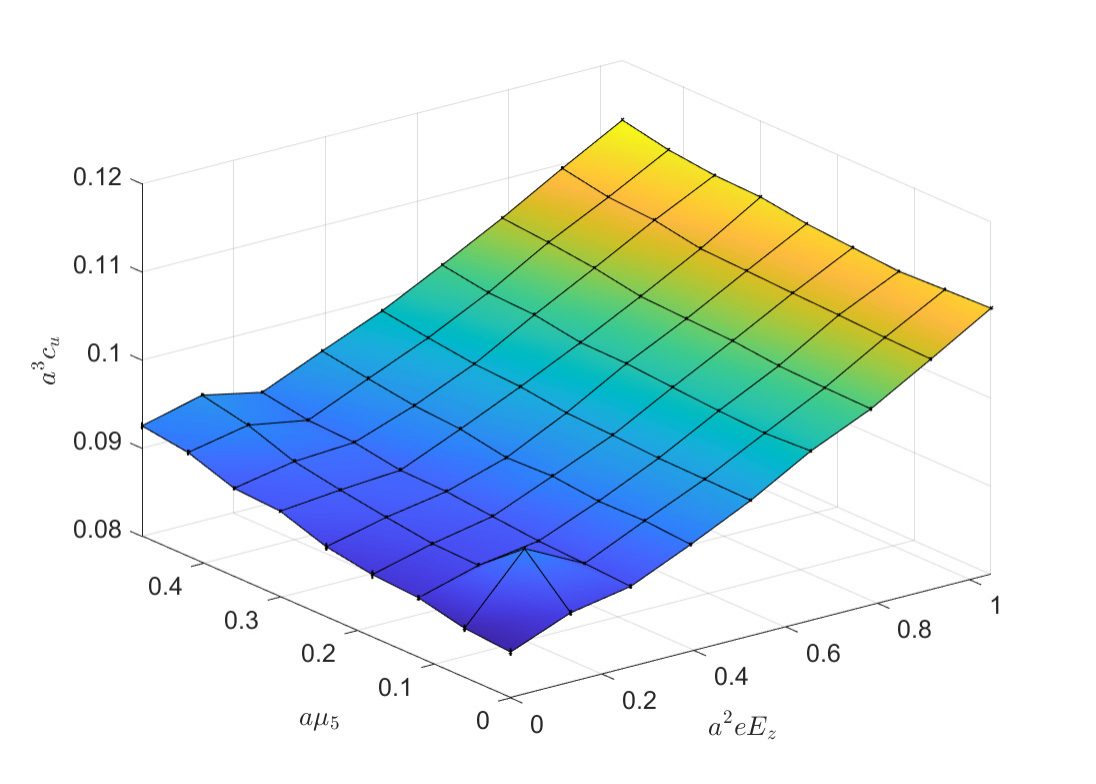}
\includegraphics[width=0.48\hsize]{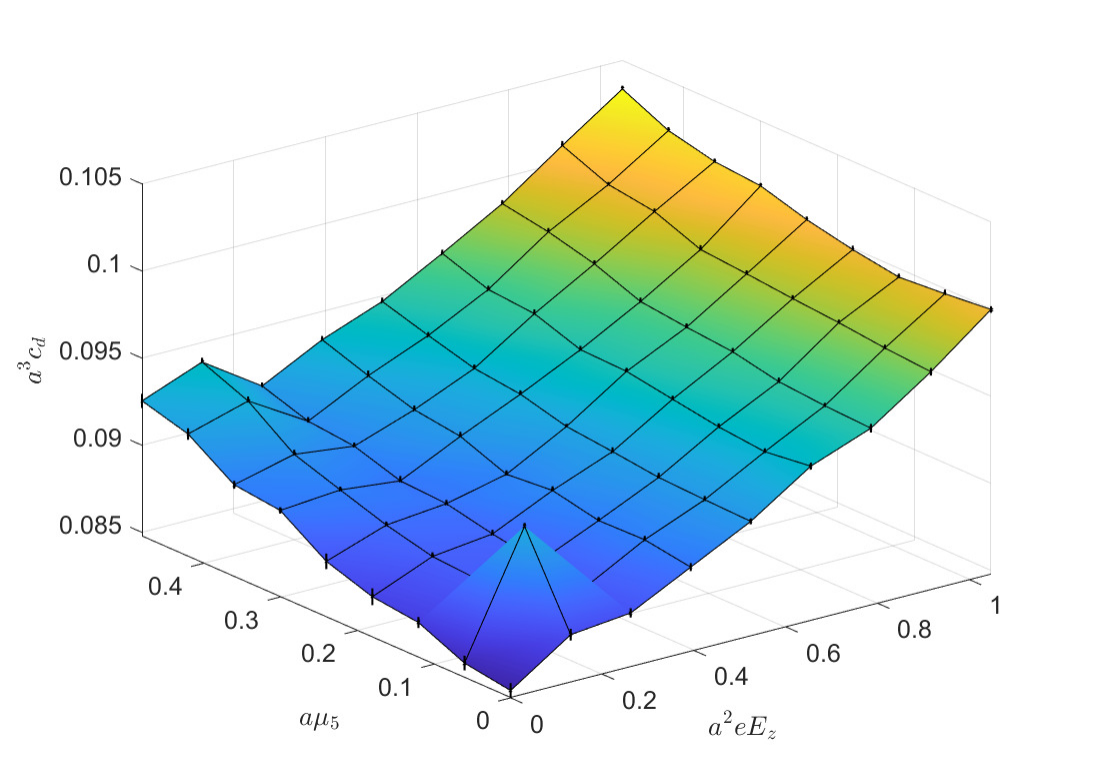}\\
\includegraphics[width=0.48\hsize]{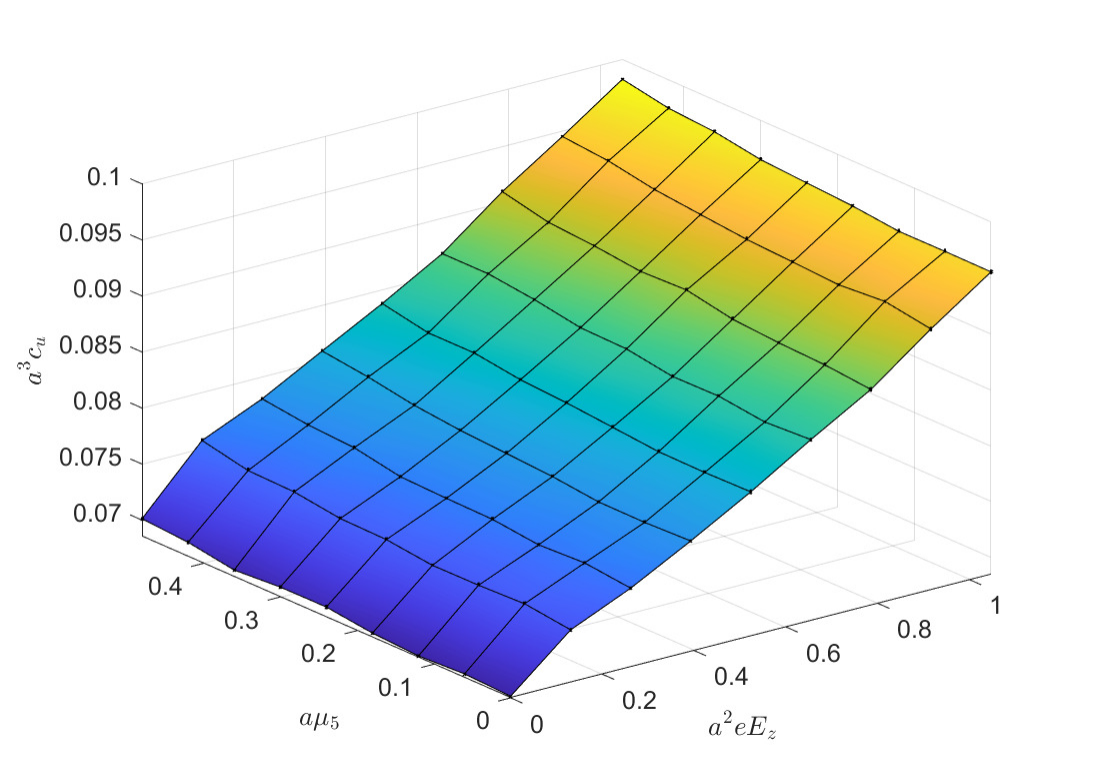}
\includegraphics[width=0.48\hsize]{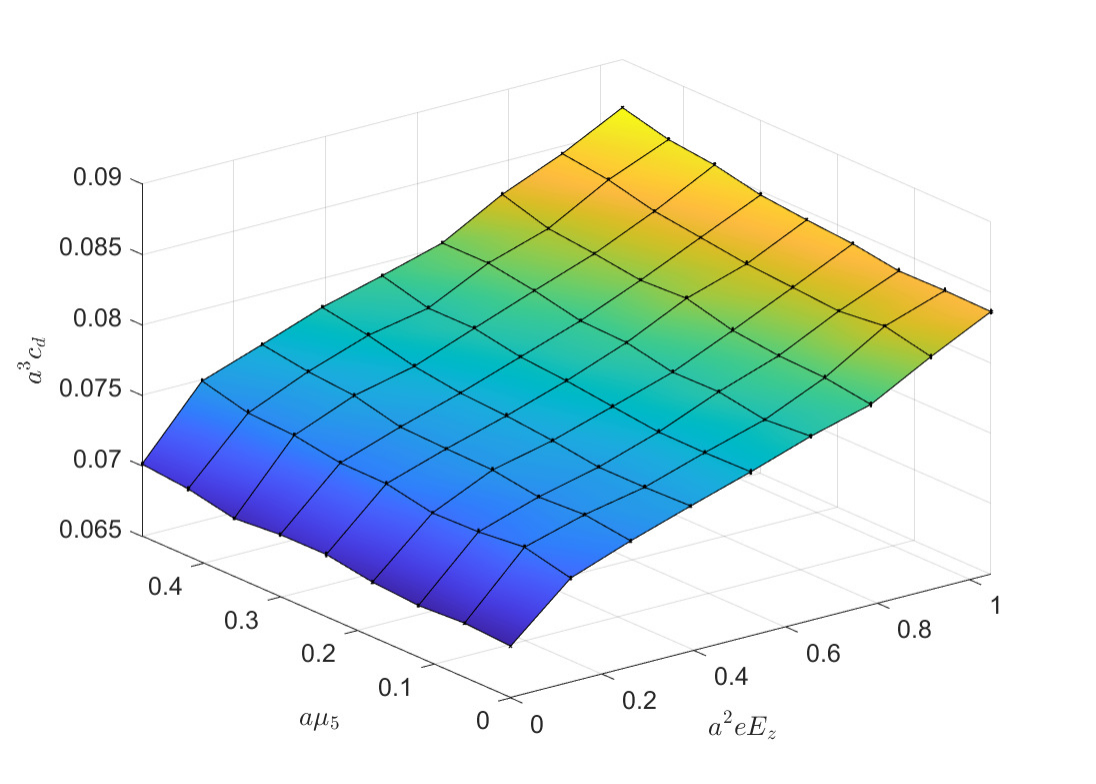}
\caption{\label{fig:chiralcondensation}
Chiral condensation as a function of $a^eE_z$ and $\mu _5$ at different temperatures for different quarks.
The first row is for the case of $\beta = 5.3, T=202.5\;{\rm MeV}$, where the left panel shows the $u$ quark and the right panel shows the $d$ quark.
The second row is for the case of $\beta = 5.4, T=270.5\;{\rm MeV}$, where the left panel shows the $u$ quark and the right panel shows the $d$ quark.}
\end{center}
\end{figure}
The chiral condensation as a function of $a^2eE_z$ and $\mu _5$ for different quarks at different temperatures are shown in Fig.~\ref{fig:chiralcondensation}.
It can be shown that, the imaginary electric field breaks the chiral symmetry.
The behavior indicates that, after Wick rotation back to the case of real electric field, at a small $a^2eE_z$, the real electric field will restore the chiral symmetry.
Apart from that, it is also shown that the chiral chemical potential breaks the chiral symmetry.
Both the results consistent with the theoretical predictions.
The results for real electric field will be presented in next subsection.
However, in the case of the $\beta = 5.3, T=202.5\;{\rm MeV}$, a non-monotonic behavior is found at small chiral chemical potential $a\mu_5=0.06$ and electric field strength $a^2eE_z=\pi / 24$.
Apart from that, it can be found that, the effect of $\mu_5$ is much smaller at a higher temperature.

It is interesting to see the change of chiral condensation with different influences.
Denoting $c_q(E_z, \mu _5)$ as the chiral condensation at a certain strengths of electric field and chiral chemical potential, then we use,
\begin{equation}
\begin{split}
&\Delta c_q^E (\mu _5)=  c_q(a^{-2}e^{-1}\pi/3, \mu _5) - c_q(0, \mu _5),\\
&\Delta c_q^{\rm CCP} (E_z)=  c_q(E_z, 0.48 a^{-1}) - c_q(E_z, 0),
\end{split}
\label{eq.3.2}
\end{equation}
where the superscript `E' and `CCP' stand for the changes caused by electric field and chiral chemical potential, respectively.

\begin{figure}[htbp]
\begin{center}
\includegraphics[width=0.48\hsize]{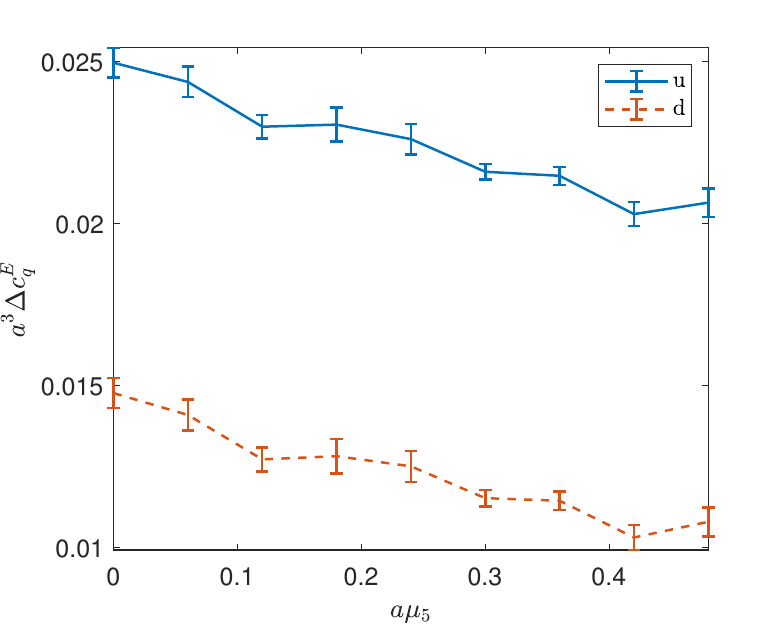}
\includegraphics[width=0.48\hsize]{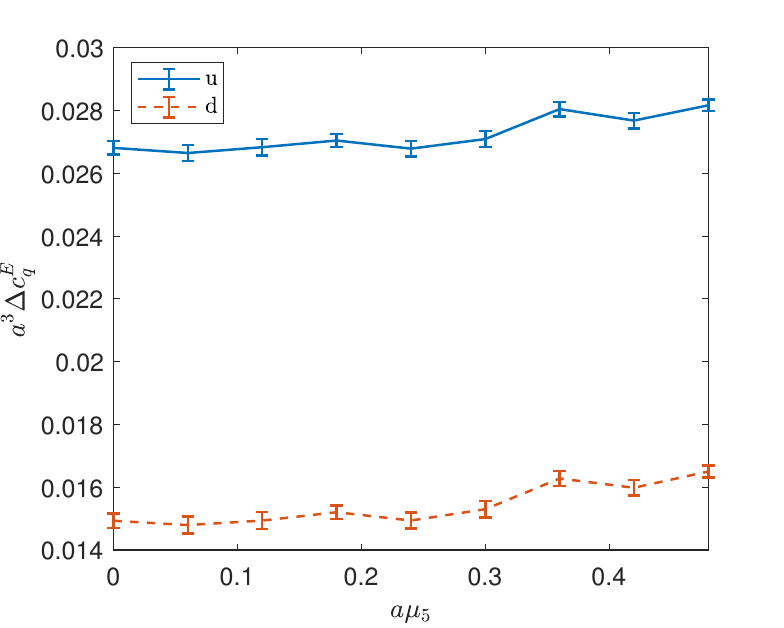}
\caption{\label{fig:deltachiralcondensationcp}
$\Delta c_q^E $ as functions of $\mu _5$ for the case of $\beta=5.3, T=202.5\;{\rm MeV}$~(the left panel) and $\beta=5.4, T=270.5\;{\rm MeV}$~(the right panel).}
\end{center}
\end{figure}
The $\Delta c_q^E$ for different temperatures are shown in Fig.~\ref{fig:deltachiralcondensationcp}.
It can be shown that, $\Delta c_u^E\approx 2\Delta c_d^E$, which is consistent to the electric charge since the $\Delta c_q^E$ is the effect of electric field on the chiral condensation at different temperatures and $\mu _5$.
The effect of the electric field is insensitive to the chiral chemical potential especially at high temperatures.
The chiral chemical potential can suppress the effect of electric field at lower temperatures, and slightly enhance the effect of electric field at high temperatures.

\begin{figure}[htbp]
\begin{center}
\includegraphics[width=0.48\hsize]{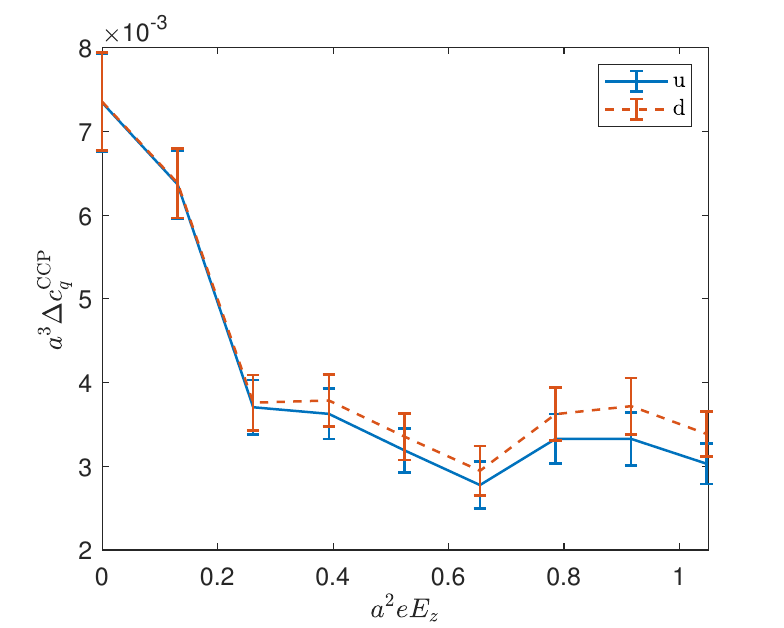}
\includegraphics[width=0.48\hsize]{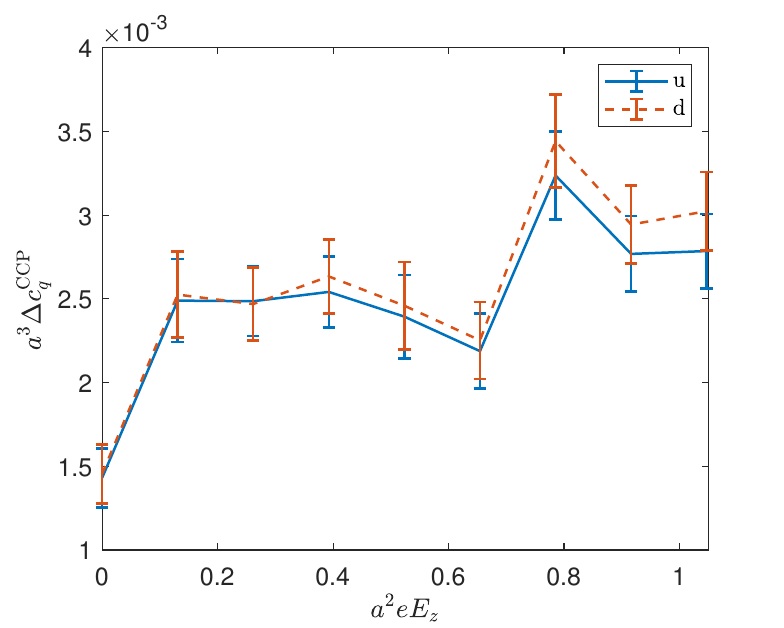}
\caption{\label{fig:deltachiralcondensatione}
$\Delta c_q^{\rm CCP} $ as functions of $\mu _5$ for the case of $\beta=5.3, T=202.5\;{\rm MeV}$~(the left panel) and $\beta=5.4, T=270.5\;{\rm MeV}$~(the right panel).}
\end{center}
\end{figure}
The $\Delta c_q^{\rm CCP}$ for different temperatures are shown in Fig.~\ref{fig:deltachiralcondensatione}.
$\Delta c_q^{\rm CCP}$ are orders of magnitudes smaller than $\Delta c_q^E$, indicating that the effect of chiral chemical potential is much smaller.
The effect of chiral chemical potential is almost the same for $u$ and $d$ quarks, however, with large electric field there is a small difference between $u$ and $d$ quarks.
Note with large electric field $c_u>c_d$, so the above feature indicates that the more severely the chiral symmetry is broken, the greater the effect of chiral chemical potential.
It can be also observed that the dependence of the effect of chiral chemical potential on the electric field is not monotonic.

In both Figs.~\ref{fig:chiralcondensation} and \ref{fig:deltachiralcondensatione}, it can be found that, at high temperatures the case of $a^2eE_z = 0$ is different from the case of $a^2eE_z\neq 0$, i.e., the change of chiral condensation implies a discontinuity at non-zero imaginary electric field, no matter whether the chiral chemical potential presents or not.
These discontinuities are discussed at length in Ref.~\cite{Endrodi:2023wwf}.

\begin{figure}[htbp]
\begin{center}
\includegraphics[width=0.48\hsize]{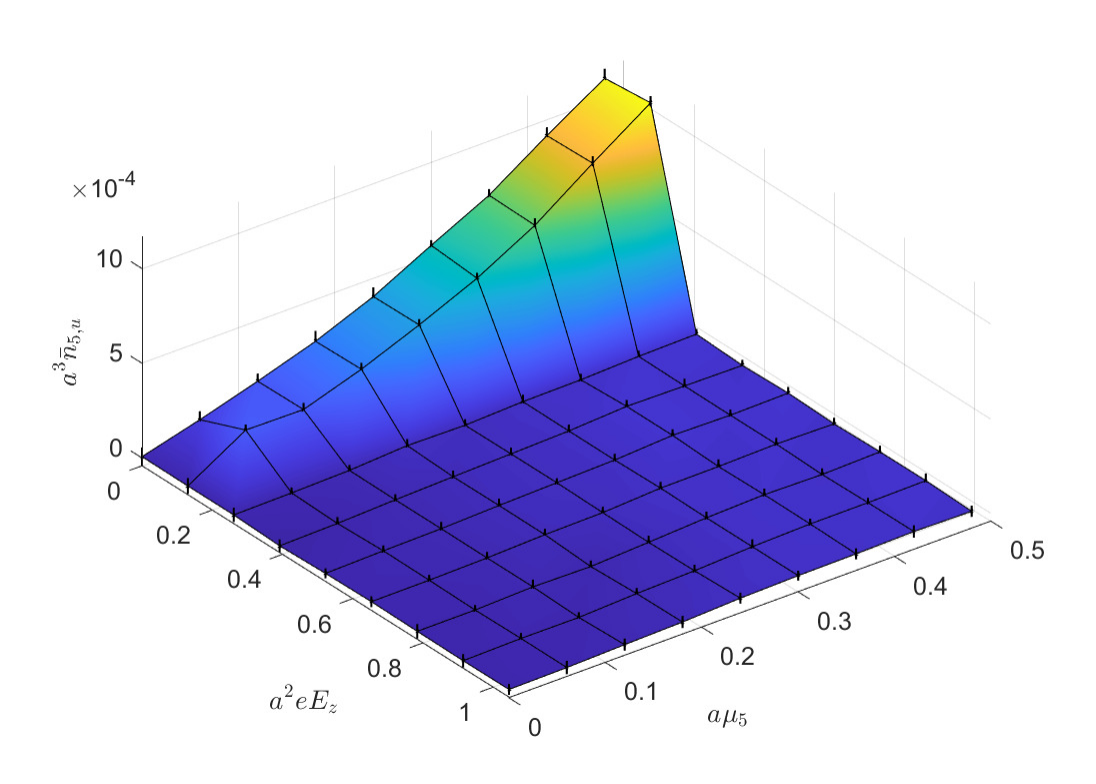}
\includegraphics[width=0.48\hsize]{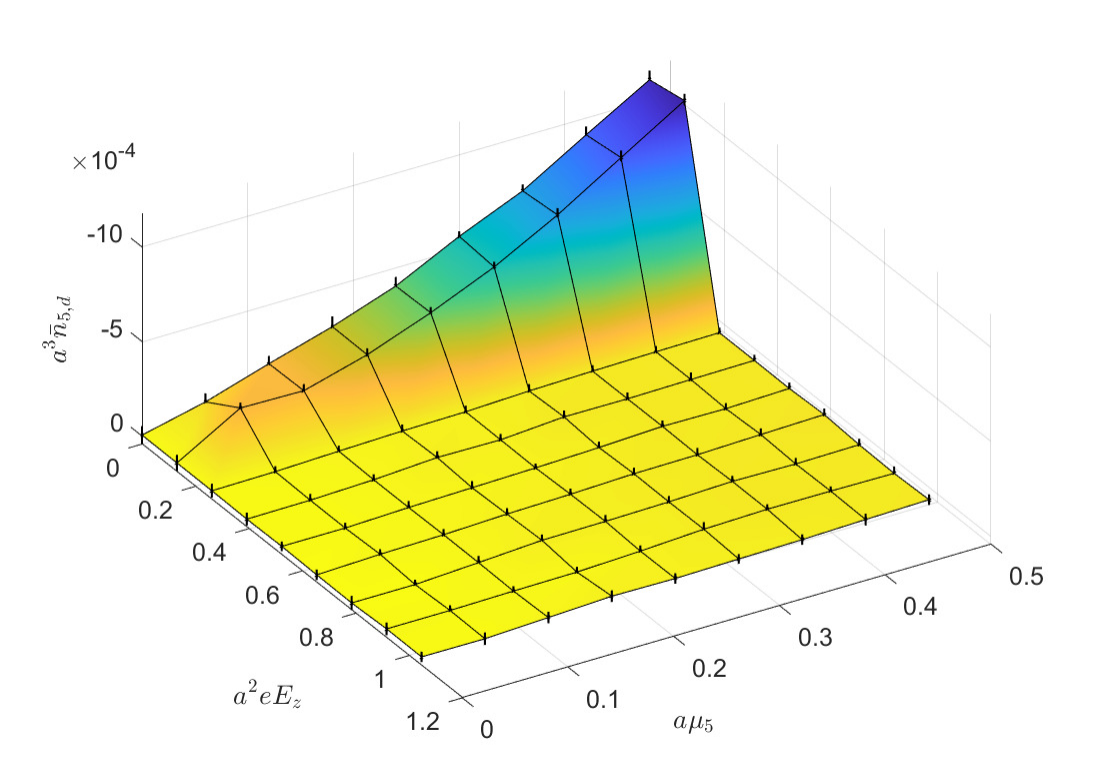}\\
\includegraphics[width=0.48\hsize]{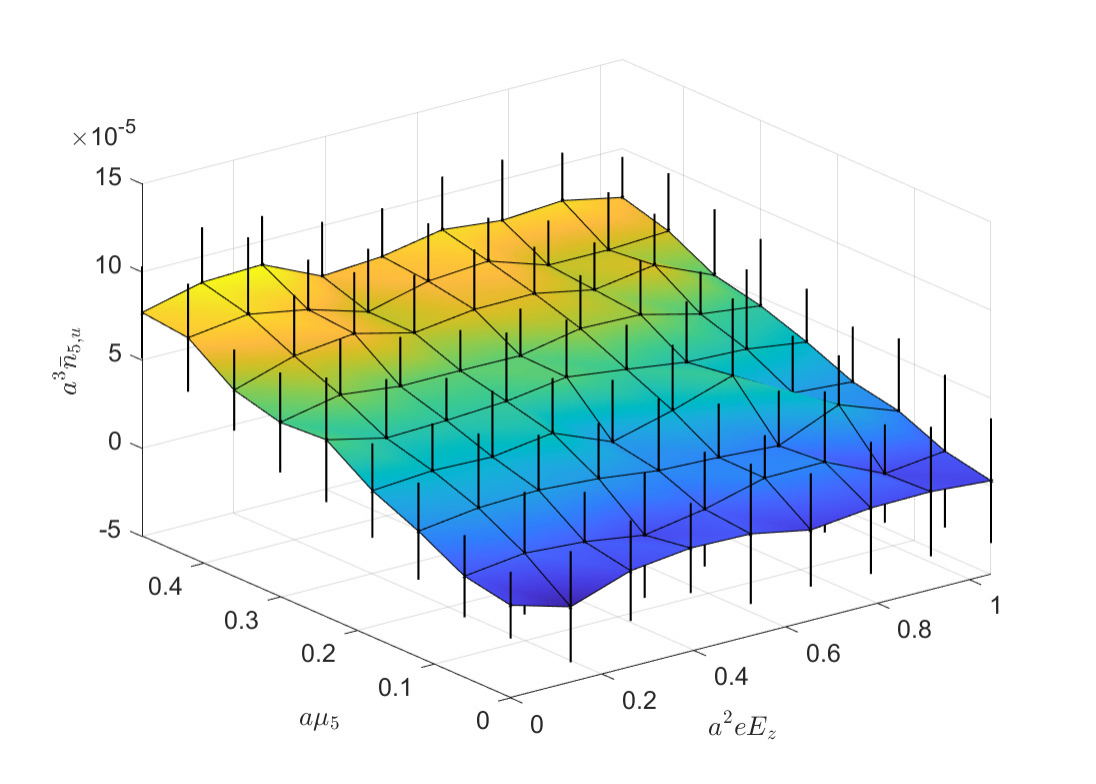}
\includegraphics[width=0.48\hsize]{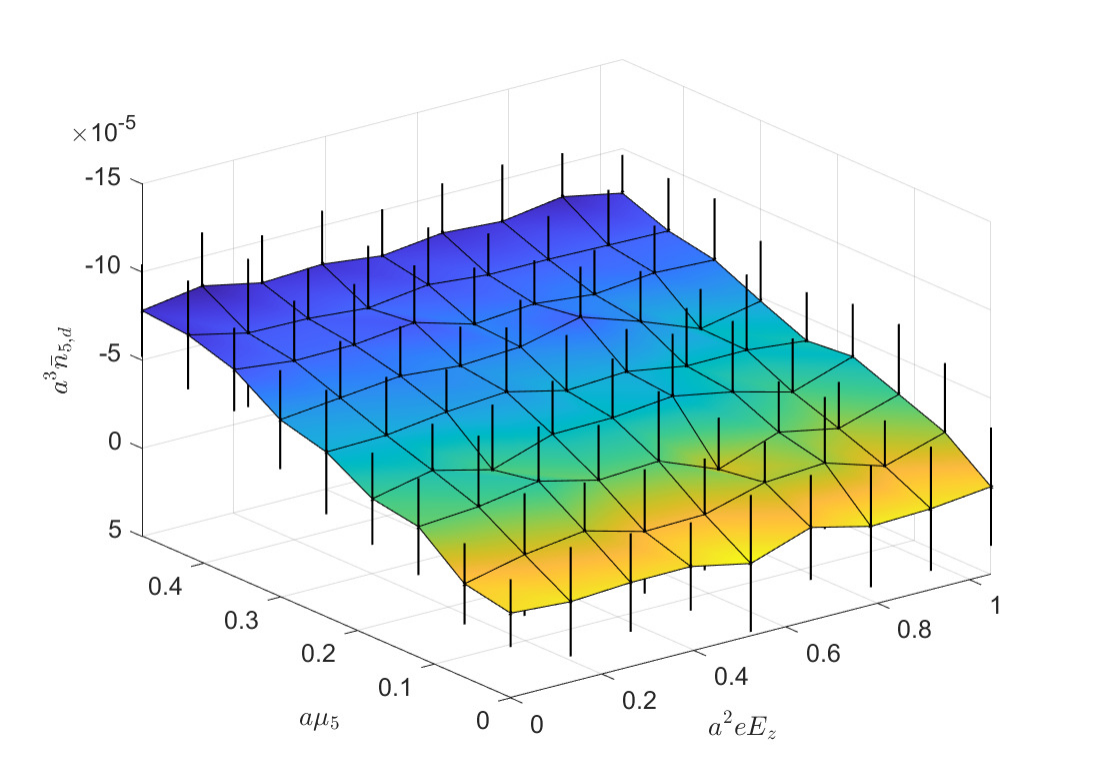}
\caption{\label{fig:n5cond}
$\bar{n}_5$ as functions of $E_z$ and $\mu _5$ at different temperatures and for different quarks.
The first row is for the case of $\beta = 5.3, T=202.5\;{\rm MeV}$, where the left panel shows the $u$ quark and the right panel shows the $d$ quark.
The second row is for the case of $\beta = 5.4, T=270.5\;{\rm MeV}$, where the left panel shows the $u$ quark and the right panel shows the $d$ quark.}
\end{center}
\end{figure}

\begin{table*}[hbtp]
\centering
\begin{tabular}{c|c|c|c|c|c|c|c|c|c}
\hline
\multirow{2}{*}{$n$} & \multicolumn{8}{c}{$k$} \\
\cline{2-10} 
 & 0 & 1 & 2 & 3 & 4 & 5 & 6 & 7 & 8 \\
\hline
0 & $51.2$ & $45.4$ & $36.9$ & $64.5$ & $49.9$ & $15.7$ & $23.2$ & $39.4$ & $46.8$ \\
1 & $39.4$ & $17.7$ & $26.4$ & $24.6$ & $18.8$ & $17.1$ & $27.2$ & $28.2$ & $24.1$ \\
2 & $30.5$ & $18.1$ & $17.6$ & $25.1$ & $21.0$ & $29.2$ & $20.9$ & $21.6$ & $20.5$ \\
3 & $24.3$ & $23.3$ & $17.4$ & $22.6$ & $21.5$ & $23.4$ & $23.3$ & $30.7$ & $18.4$ \\
4 & $14.0$ & $16.9$ & $25.1$ & $25.6$ & $11.2$ & $21.2$ & $20.2$ & $18.1$ & $20.6$ \\
5 & $15.7$ & $16.0$ & $18.7$ & $19.5$ & $18.9$ & $20.7$ & $27.3$ & $21.6$ & $20.5$ \\
6 & $28.7$ & $12.7$ & $26.5$ & $18.2$ & $22.3$ & $19.5$ & $17.2$ & $20.3$ & $19.3$ \\
7 & $22.6$ & $14.7$ & $13.3$ & $15.3$ & $18.0$ & $27.6$ & $13.7$ & $11.3$ & $25.8$ \\
8 & $14.9$ & $21.8$ & $18.0$ & $18.5$ & $15.1$ & $12.2$ & $14.9$ & $18.6$ & $15.5$ \\
\hline
\end{tabular}
\caption{$\tau _{int}$ at $a^2eE_z=k\pi/24$ and $a\mu _5=0.06n$ when $\beta=5.3$.}
\label{table:tint530}
\end{table*}

\begin{table*}[hbtp]
\centering
\begin{tabular}{c|c|c|c|c|c|c|c|c|c}
\hline
\multirow{2}{*}{$n$} & \multicolumn{8}{c}{$k$} \\
\cline{2-10} 
 & 0 & 1 & 2 & 3 & 4 & 5 & 6 & 7 & 8 \\
\hline
0 & $ 9.4$ & $13.8$ & $19.4$ & $17.9$ & $32.2$ & $20.5$ & $13.9$ & $24.0$ & $17.3$ \\
1 & $23.5$ & $29.6$ & $40.4$ & $16.7$ & $27.0$ & $24.0$ & $28.9$ & $43.5$ & $22.8$ \\
2 & $19.0$ & $18.3$ & $22.9$ & $32.2$ & $13.3$ & $18.1$ & $27.2$ & $19.0$ & $17.1$ \\
3 & $15.0$ & $28.5$ & $21.7$ & $21.1$ & $31.8$ & $11.7$ & $24.5$ & $28.9$ & $20.5$ \\
4 & $33.8$ & $22.9$ & $34.2$ & $25.5$ & $22.7$ & $15.5$ & $23.6$ & $22.9$ & $15.1$ \\
5 & $20.5$ & $11.7$ & $18.5$ & $27.8$ & $12.9$ & $17.4$ & $13.6$ & $12.1$ & $17.3$ \\
6 & $26.5$ & $31.0$ & $18.9$ & $34.7$ & $23.2$ & $17.8$ & $13.5$ & $12.7$ & $22.3$ \\
7 & $25.3$ & $13.7$ & $14.6$ & $20.1$ & $37.1$ & $10.6$ & $15.6$ & $19.8$ & $12.9$ \\
8 & $23.7$ & $32.7$ & $31.6$ & $17.8$ & $15.9$ & $25.2$ & $25.0$ & $18.6$ & $ 9.3$ \\
\hline
\end{tabular}
\caption{Same as Table~\ref{table:tint530} but for $\beta=5.4$.}
\label{table:tint540}
\end{table*}

\begin{figure*}[htbp]
\begin{center}
\includegraphics[width=0.98\hsize]{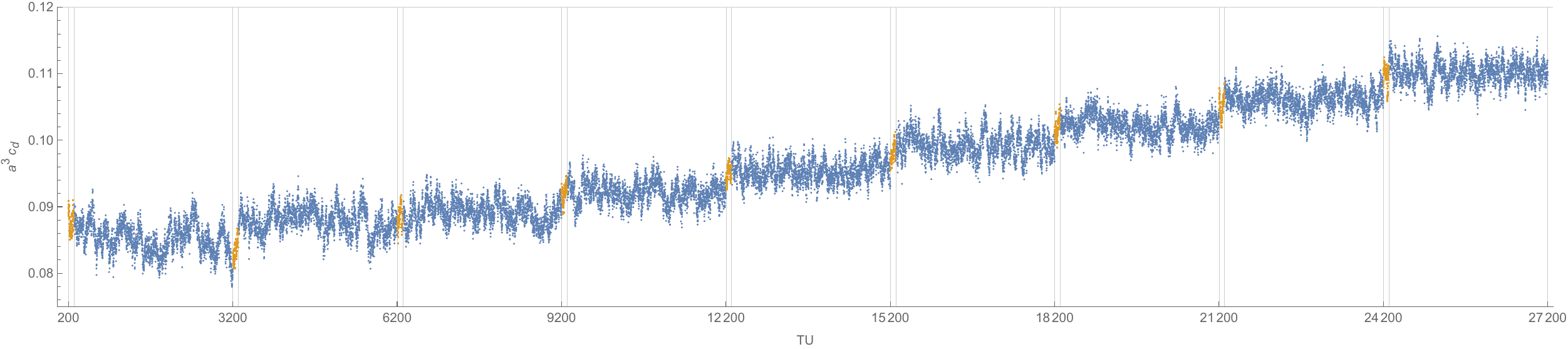}\\
\includegraphics[width=0.98\hsize]{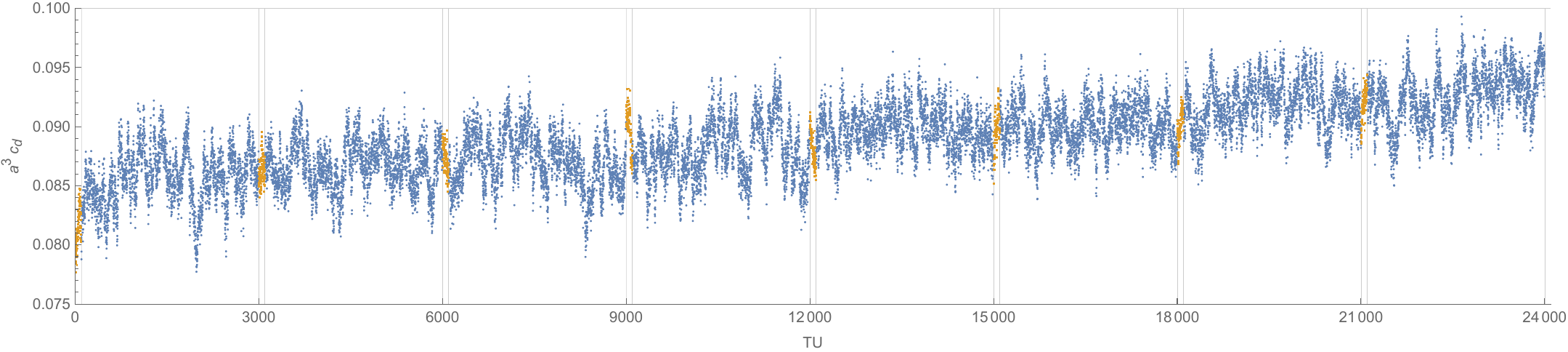}\\
\caption{\label{fig:chiralhistory}
The history of chiral condensation with growing TU in the case of $\beta = 5.3$ and $\mu_5=0$ with growing $E_z$~(the top panel), and the case of $\beta = 5.3$ and $E_z=0$ with growing $\mu_5$~(the bottom panel).
The first $200$ configurations of the case $\beta = 5.3$ and $\mu_5=0$ are discarded and therefore not shown.
The configurations correspond to thermalization are marked as yellow.
}
\end{center}
\end{figure*}
Only $100$ configurations are discarded seems too little. 
However, on the one hand we incrementally increase the electric field strength/chiral chemical potential, i.e., when using new electric field strengths and chiral chemical potential, we change from the previous value and start the simulation from the configuration at the previous value, as shown in Fig.~\ref{fig:simulationpath}. 
On the other hand, the $8\times8\times 24\times 6$ lattice is a relatively small lattice with a relatively large fluctuation, which also makes it easier to reach equilibrium. 
Taking the case of $\beta = 5.3$ and $\mu_5=0$ with growing $E_z$, and the case of $\beta = 5.3$ and $E_z=0$ with growing $\mu_5$ as examples, the history of chiral condensation with TU are shown in Fig.~\ref{fig:chiralhistory}. 
In addition, the results for $\tau _{int}$ in Tables~\ref{table:tint530} and \ref{table:tint540}.  

\subsection{\label{sec:3.3}Analytical continuation of chiral condensation}

Since the imaginary electric field is used, to compare with the real physics, one needs to Wick rotate the electric back to a real electric field.
The result of analytical continuation depends the ansatz to used.
Following the study of the imaginary chemical potential, we use the following three different ansatzes and take their differences as systematic errors~\cite{Bellwied:2016gtm}.
We assume that, at a given $\mu _5$, the chiral condensation is a function of $E_z$,
\begin{equation}
\begin{split}
&f_1(E_z) = \alpha+\beta (a^2eE_z)^2 + \rho (a^2eE_z)^4,\\
&f_2(E_z) = \frac{ \alpha+\beta (a^2eE_z)^2}{ 1+\rho (a^2eE_z)^2},\\
&f_3(E_z) = \alpha+\beta (a^2eE_z)^2 + \rho \frac{\sin (a^2eE_z)}{a^2eE_z},\\
\end{split}
\label{eq.3.chiralansatz}
\end{equation}
where $\alpha$, $\beta$ and $\rho$ are parameters to be fitted.

Due to the non-monotonic feature of the chiral condensation, the chiral condensations at $a^2eE_z=\pi/24$ and $\pi/12$ are ignored.
The fit of the results are shown in the \ref{sec:ap1}.
The non-monotonic of the chiral condensation appears as an oscillation damped by a large imaginary electric field.

\begin{figure}[htbp]
\begin{center}
\includegraphics[width=0.48\hsize]{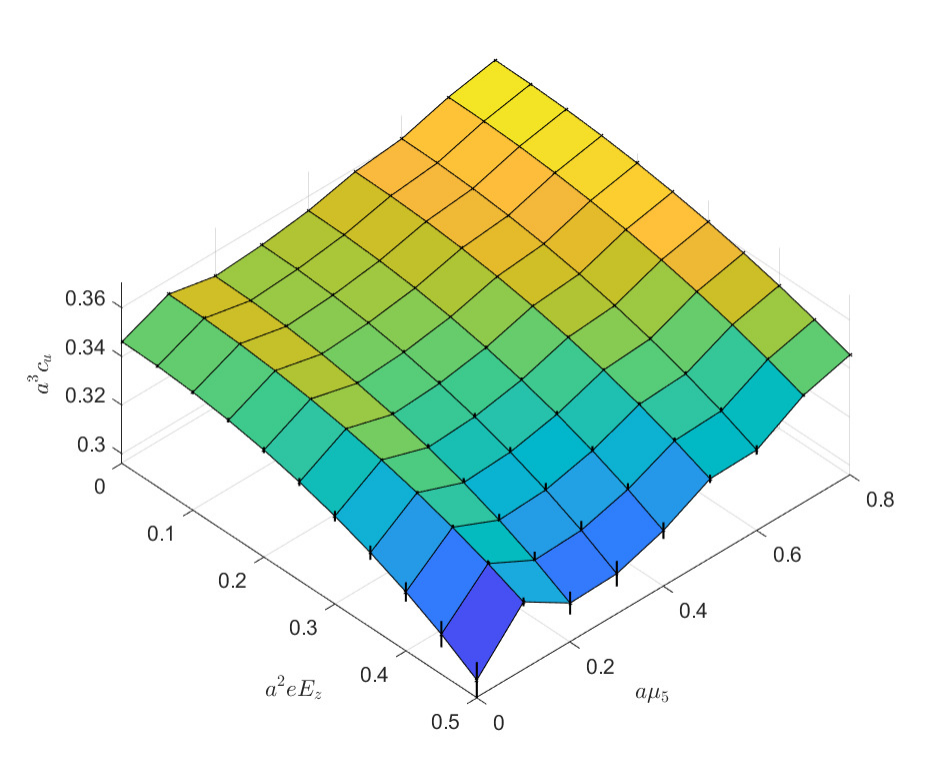}
\includegraphics[width=0.48\hsize]{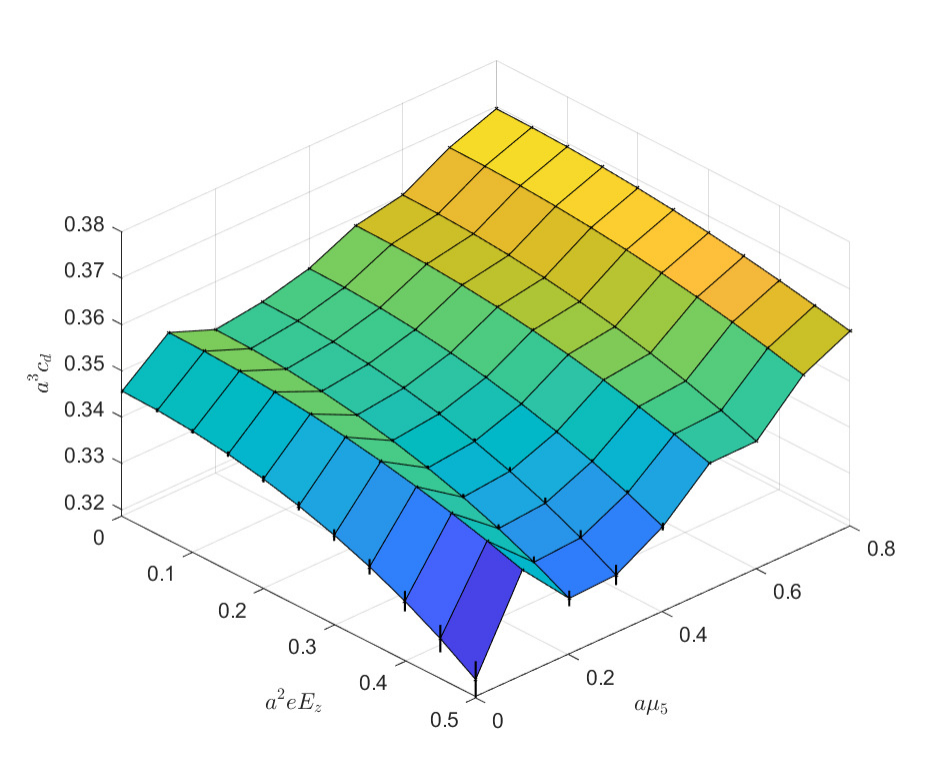}\\
\includegraphics[width=0.48\hsize]{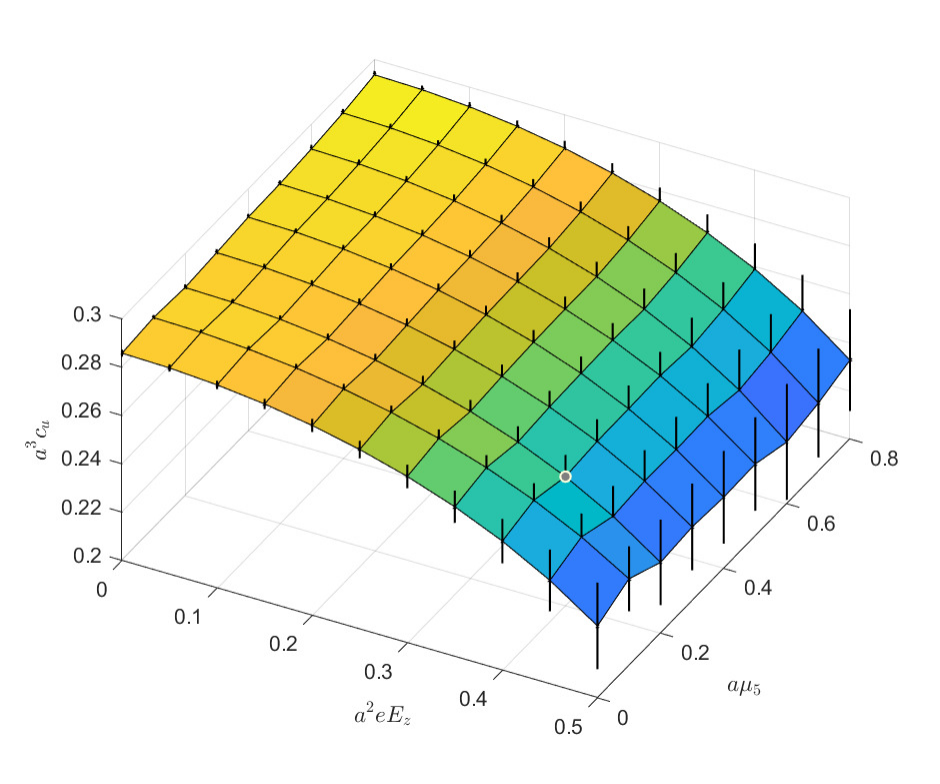}
\includegraphics[width=0.48\hsize]{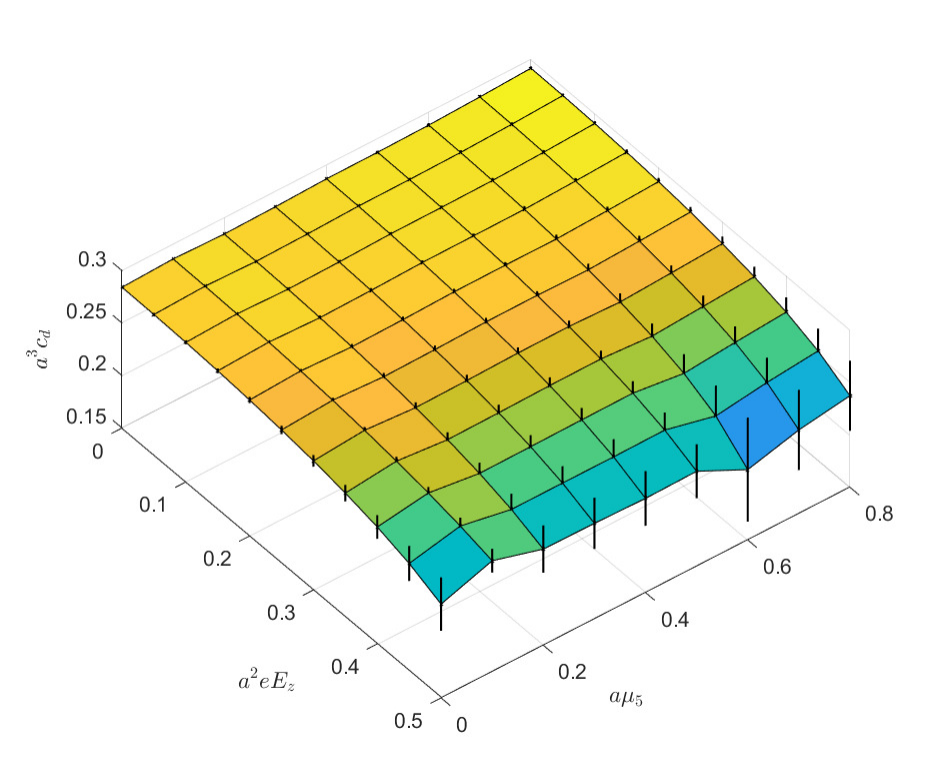}
\caption{\label{fig:realelectric}
Chiral condensations as functions of real electric field strength and $\mu _5$.
The value is taken as medians among different ansatzes, and the error bars represent the maximums and minimums of different ansatzes.
The first row is for the case of $\beta = 5.3, T=202.5\;{\rm MeV}$, where the left panel shows the case of $c_u$ and the right panel shows the case of $c_d$.
The second row is for the case of $\beta = 5.4, T=270.5\;{\rm MeV}$, where the left panel shows the case of $c_u$ and the right panel shows the case of $c_d$.}
\end{center}
\end{figure}
After the electric field is Wick rotated back to the real electric field, the chiral condensation are shown in Fig.~\ref{fig:realelectric}.
For larger real $a^2eE_z$, the systematic error is large, therefore only chiral condensations for $a^2eE_z < 0.5$ are shown.
The results indicate that the real electric field will restore the chiral symmetry which is consistent with the theoretical predictions.
Meanwhile, a large chiral chemical potential breaks the chiral symmetry which is also consistent with the theoretical predictions.
However, for a small chiral chemical potential, the non-monotonic feature still exists.

\subsection{\label{sec:3.4}Charge density}

Another quantity of interest is the charge density $\bar{n}_5$.
$\bar{n}_5$ as a function of $a^2eE_z$ and $a\mu _5$ for different quarks at different temperatures are shown in Fig.~\ref{fig:n5cond}.
The results indicates that $\bar{n}_{5,u}>0$, $n_{5,d}<0$ and $\bar{n}_{5,u}+\bar{n}_{5,d}\approx 0$.
It can be seen that $\bar{n}_5$ is sensitive to $\mu _5$ at low temperatures.
This phenomenon is compatible with the previous one, since the more severely the chiral symmetry is broken, the greater the effect of the chiral chemical potential.
However, we also note that, the $\bar{n}_5$ is only sensitive to $\mu _5$ with weak electric fields.
Another interesting phenomenon is that, at $a^2eE_z=\pi/24$ and $\beta= 5.3$, $\bar{n}_5$ is not a monotonic function of $\mu _5$.
This phenomenon matches the behavior of chiral condensation.

\subsection{\label{sec:3.5}Polyakov loop}

It has been found in Ref.~\cite{Yang:2022zob} that, there is R-W transition when imaginary electric field is applied, and the Polyakov loop oscillates along the ${\bf z}$-axis. 
In this paper, we measure the $\langle P(z)\rangle$, where $P(z)$ is defined as,
\begin{equation}
\begin{split}
P(z) = \frac{1}{L_xL_y}\sum _{n_x,n_y}\prod _{n_{\tau}}U_{\tau}\left(n=\left(n_x,n_y,a^{-1}z,n_{\tau}\right)\right),\\
\end{split}
\label{eq.3.3}
\end{equation}
where the product over $n_{\tau}$ is ordered in the $\tau$ direction.

It was pointed out that, the Polyakov loop as a function of $z$ can be fitted using $\langle P(z)\rangle = A_p + \sum _{q=u,d} C_q \exp(L_{\tau}Q_qiazeE_z)$, where $A_p$ is a complex number, $C_q$ are real numbers depending on temperature.
In this paper, we find that the ansatz can be further simplified and one can remove another degree of freedom to use 
\begin{equation}
\begin{split}
\langle P(z)\rangle = A_p + B_p\sum _{q=u,d} \left|Q_q\right| \exp(L_{\tau}Q_qiazeE_z).
\end{split}
\label{eq.3.4}
\end{equation}
where $A_p$ is a complex number, $B_p$ is a real number, both depending on temperature and the chiral chemical potential.

The oscillation of $\langle P \rangle$ for the case with smaller electric field is more significant, therefore we focus on the cases that $a^2eE_z=\pi / 24$ and $a^2eE_z=\pi / 12$. 
For different chiral chemical potential, the results at $\beta=5.3~(T=202.5\;{\rm MeV})$ and $\beta=5.4~(T=270.5\;{\rm MeV})$ are shown in \ref{sec:ap2}.

\begin{figure}[htbp]
\begin{center}
\includegraphics[width=0.48\hsize]{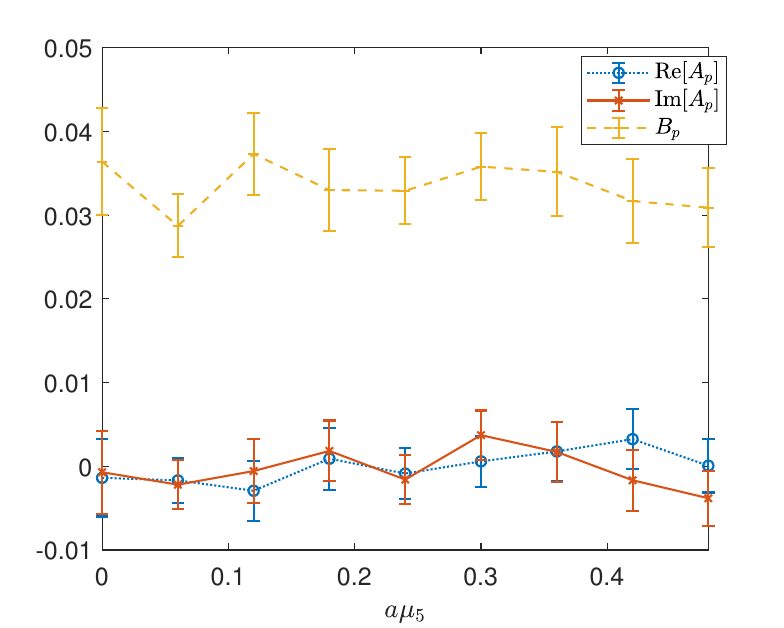}
\includegraphics[width=0.48\hsize]{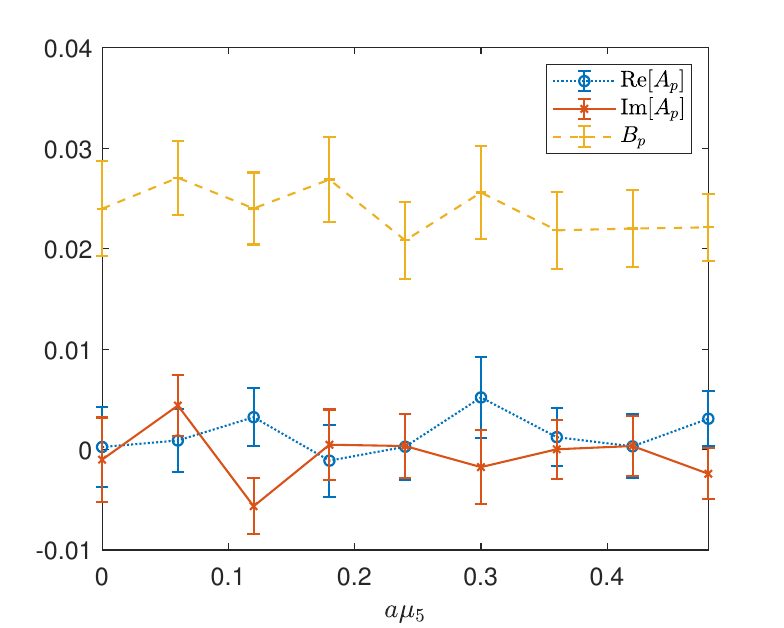}\\
\includegraphics[width=0.48\hsize]{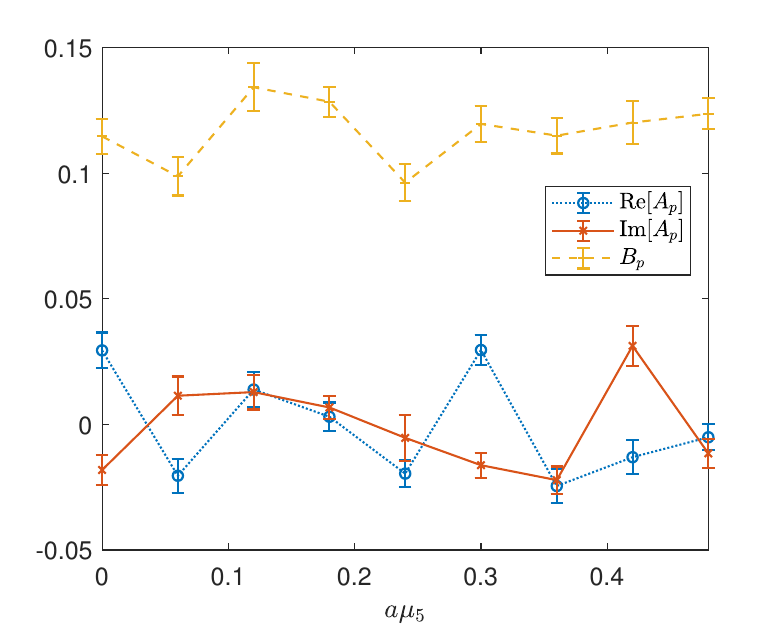}
\includegraphics[width=0.48\hsize]{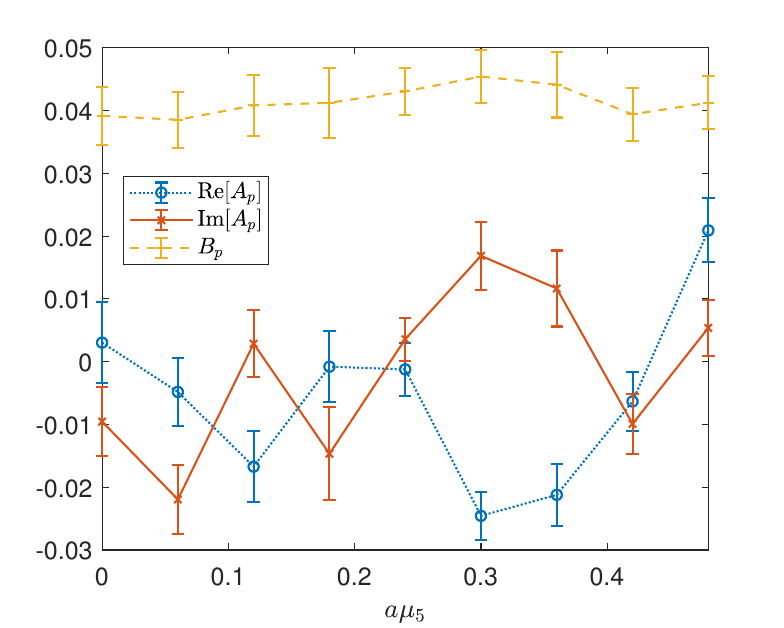}
\caption{\label{fig:polyakovfit}
$A_p$ and $B_p$ in Eq.~(\ref{eq.3.4}) as functions of the chiral chemical potential at different temperatures and electric fields.
The first row is for the case of $\beta = 5.3, T=202.5\;{\rm MeV}$, where the left panel shows the case of $a^2eE_z=\pi/24$ and the right panel shows the case of $a^2eE_z=\pi/12$.
The second row is for the case of $\beta = 5.4, T=270.5\;{\rm MeV}$, where the left panel shows the case of $a^2eE_z=\pi/24$ and the right panel shows the case of $a^2eE_z=\pi/12$.}
\end{center}
\end{figure}
It can be seen from Fig.~\ref{fig:polyakovfit} that, the behavior is insensitive to the chiral chemical potential, at least in the ranges of parameters investigated in this paper.
Compared with $B_p$, the $A_p$ is negligible because $|A_p|\ll B_p$, and therefore the winding numbers of the Polyakov loops $P(z)$ around the origin of the complex plane are none-zero.
Meanwhile, the dependency of $A_p$ and $B_p$ on chiral chemical potential is negligible.

The effect of the chiral chemical potential on the Polyakov loop is also studied in Refs.~\cite{Braguta:2015zta,Kotov:2015hxr}.
In this paper, the case of $\beta=5.3$ corresponds to $T=202.5$ and $\mu _5 \leq 583.2\;{\rm MeV}$, the case of $\beta=5.4$ corresponds to $T=270.5$ and $\mu _5 \leq 779.0\;{\rm MeV}$, respectively.
For the parameters in this range, the effect of the chiral chemical potential on the Polyakov loop is small, however, it can be expected that the effect of the chiral chemical potential is much more significant for $\mu _5\sim \;1\;{\rm GeV}$~\cite{Braguta:2015zta,Kotov:2015hxr}.

\section{\label{sec:4}Summary}

For theoretical and experimental reasons, the case of simultaneous chiral chemical potential and external electric field is a scenario worth investigating. 
In this paper, this case is studied using lattice QCD approach with $N_f = 1+1$ dynamic staggered fermions. 
The electric field is introduced as an imaginary electric field to avoid the sign problem.

It is confirmed that, generally, the chiral chemical potential and the imaginary electric field break the chiral symmetry, which is consistent with the results obtained by various theoretical studies.
However, in the case of the $\beta = 5.3, T=202.5\;{\rm MeV}$, a non-monotonic behavior is observed at small chiral chemical potential $a\mu_5=0.06$ and electric field strength $a^2eE_z=\pi / 24$.
It is also found that,  the chiral chemical potential can suppress the effect of electric field at lower temperatures, and slightly enhance the effect of electric field at high temperatures.
The dependence of $n_5$ on $a^2eE_z$ and $\mu _5$ is also investigated.

The Polyakov loop is also studied.
It is found that, the behavior of Polyakov is barely affected by the chiral chemical potential.
This should be due to the range of parameters taken in this paper (i.e. temperature and $\mu _5$).
Besides, we find a simplification of the ansatz for the Polyakov loop used in Ref.~\cite{Yang:2022zob}, when the electric field strength is small, the ansatz $\langle P(z)\rangle = A_p + B_p\sum _{q=u,d} \left|Q_q\right| \exp(L_{\tau}Q_qiazeE_z)$ fits well.

\begin{acknowledgement}
This work was supported in part by the National Natural Science Foundation of China under Grants No. 12147214, and the Natural Science Foundation of the Liaoning Scientific Committee No.~LJKZ0978.
\end{acknowledgement}

\appendix

\section{\label{sec:ap0}A brief introduction of the autocorrelation method used in this paper}

We briefly introduce the calculation of $\tau _{int}$ in this paper.
The case of this paper corresponds to the case of only one repetition of trajectory in Ref.~\cite{Wolff:2003sm}.
Assuming $\{\alpha _i\}$ is a sequence, autocorrelation function can be defined as,
\begin{equation}
\begin{split}
&\Gamma _{\alpha }(t)=\frac{1}{N-t}\sum _{i=1}^{N-t}\left(\alpha ^i - \bar{\alpha}\right)\left(\alpha ^{i+t} - \bar{\alpha}\right),
\end{split}
\label{eq.a.1.1}
\end{equation}
where $N$ is the length of the sequence, $\bar{\alpha}$ is the mean value of $\alpha ^i$.

Assuming that $a_j^i$ are the outcomes of the measurements with the $i$-th configuration, $b=f(a_j)$ is the observable to be calculated, a sequence $\{b^i\}$ can be obtained as,
\begin{equation}
\begin{split}
&b^i=\sum _j f^i_ja_j^i,
\end{split}
\label{eq.a.1.2}
\end{equation}
with,
\begin{equation}
\begin{split}
&h_j=\sqrt{\frac{\Gamma _{a_j}(0)}{N}},\\
&f^i_j=\frac{f(a_1^i, a_2^i,\ldots,a_j^i+h_j,\ldots)-f(a_1^i, a_2^i,\ldots,a_j^i-h_j,\ldots)}{2h_j}.
\end{split}
\label{eq.a.1.3}
\end{equation}
For a positive integer $W$, defining,
\begin{equation}
\begin{split}
&c(W)=2\left(1+\frac{2W+1}{N}\right)\sum _{t=1}^W\Gamma _b(t),\\    
&\tau (W)=\frac{S}{\log\left(\frac{c(W)+2\Gamma _b(0)}{c(W)}\right)},\\
\end{split}
\label{eq.a.1.4}
\end{equation}
where $S$ is a tunable factor and a reasonable choice is $1<S<2$~\cite{Wolff:2003sm}, and then defining,
\begin{equation}
\begin{split}
&g (W)=\exp\left(-\frac{W}{\tau (W)}\right)-\frac{\tau (W)}{\sqrt{WN}},\\
\end{split}
\label{eq.a.1.5}
\end{equation}
for $W=1,2,\ldots$, the first value of $W$ such that $g(W)<0$ is chosen as an optimal $W$, which is denoted as $W_0$, than the $\tau _{int}$ can be calculated as,
\begin{equation}
\begin{split}
&\tau _{int}=\frac{\Gamma _b(0) + c({W_0})}{2\Gamma _b(0)}.\\
\end{split}
\label{eq.a.1.6}
\end{equation}

\section{\label{sec:ap1}Fit of the chiral condensation according to different ansatzes}

\begin{figure*}[htbp]
\begin{center}
\includegraphics[width=0.98\hsize]{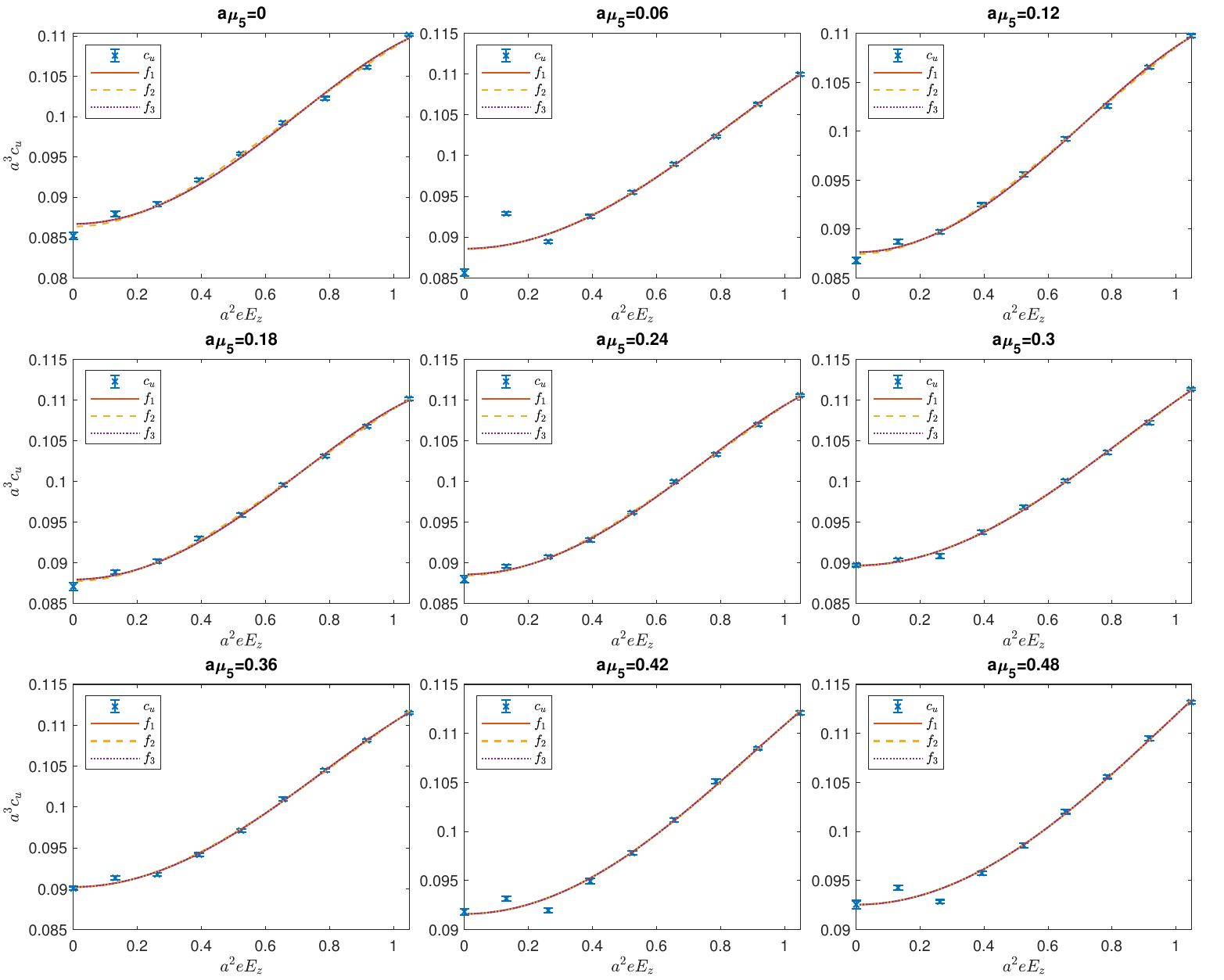}
\caption{\label{fig:chiralfit530u}
$c_u$ as a function of $a^2eE_z$ fitted with three different ansatzes in Eq.~(\ref{eq.3.chiralansatz}) at $\beta=5.3, T=202.5\;{\rm MeV}$ with different chiral chemical potentials.}
\end{center}
\end{figure*}
\begin{figure*}[htbp]
\begin{center}
\includegraphics[width=0.98\hsize]{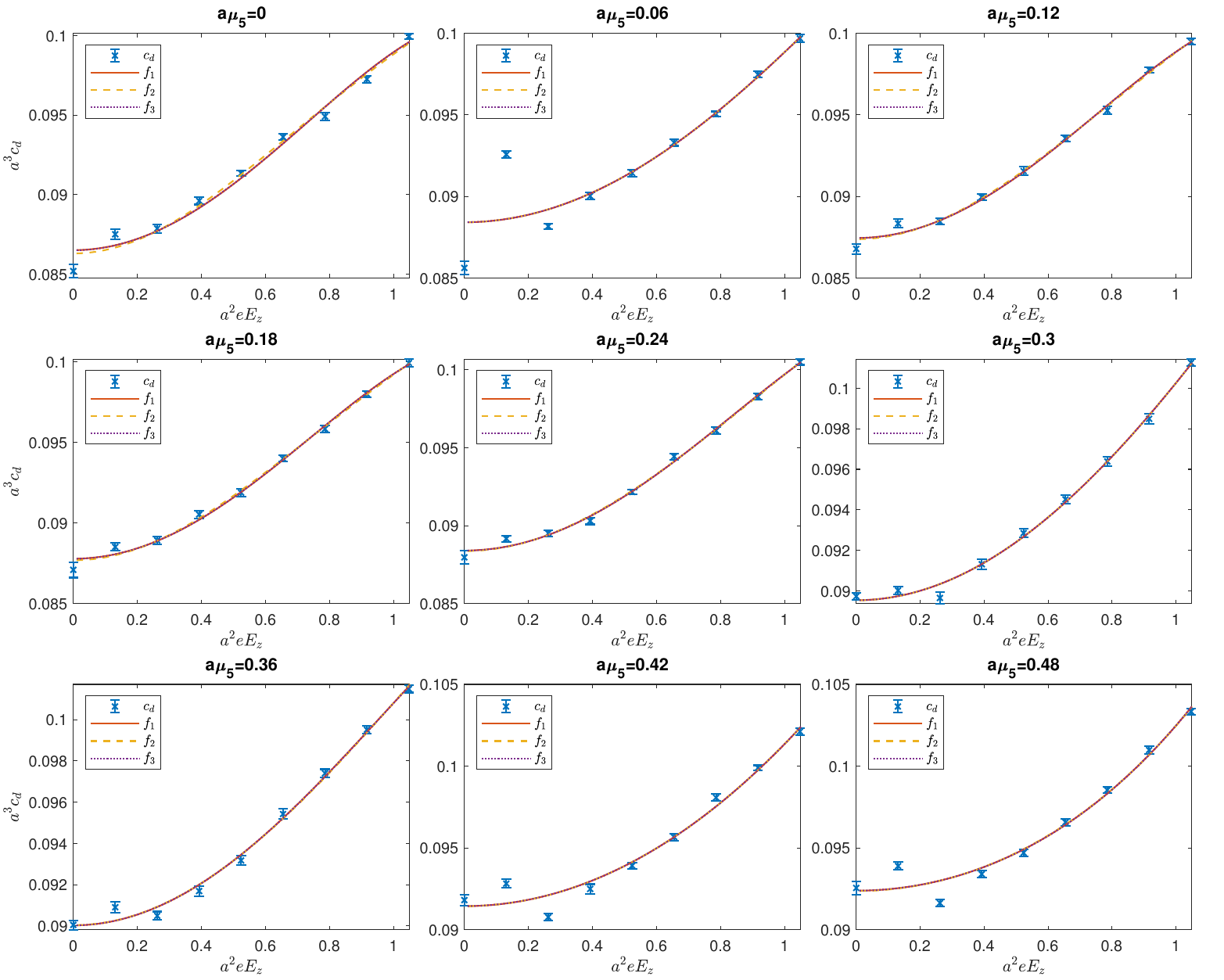}
\caption{\label{fig:chiralfit530d}
Same as Fig.~\ref{fig:chiralfit530u} but for $c_d$.}
\end{center}
\end{figure*}
\begin{figure*}[htbp]
\begin{center}
\includegraphics[width=0.98\hsize]{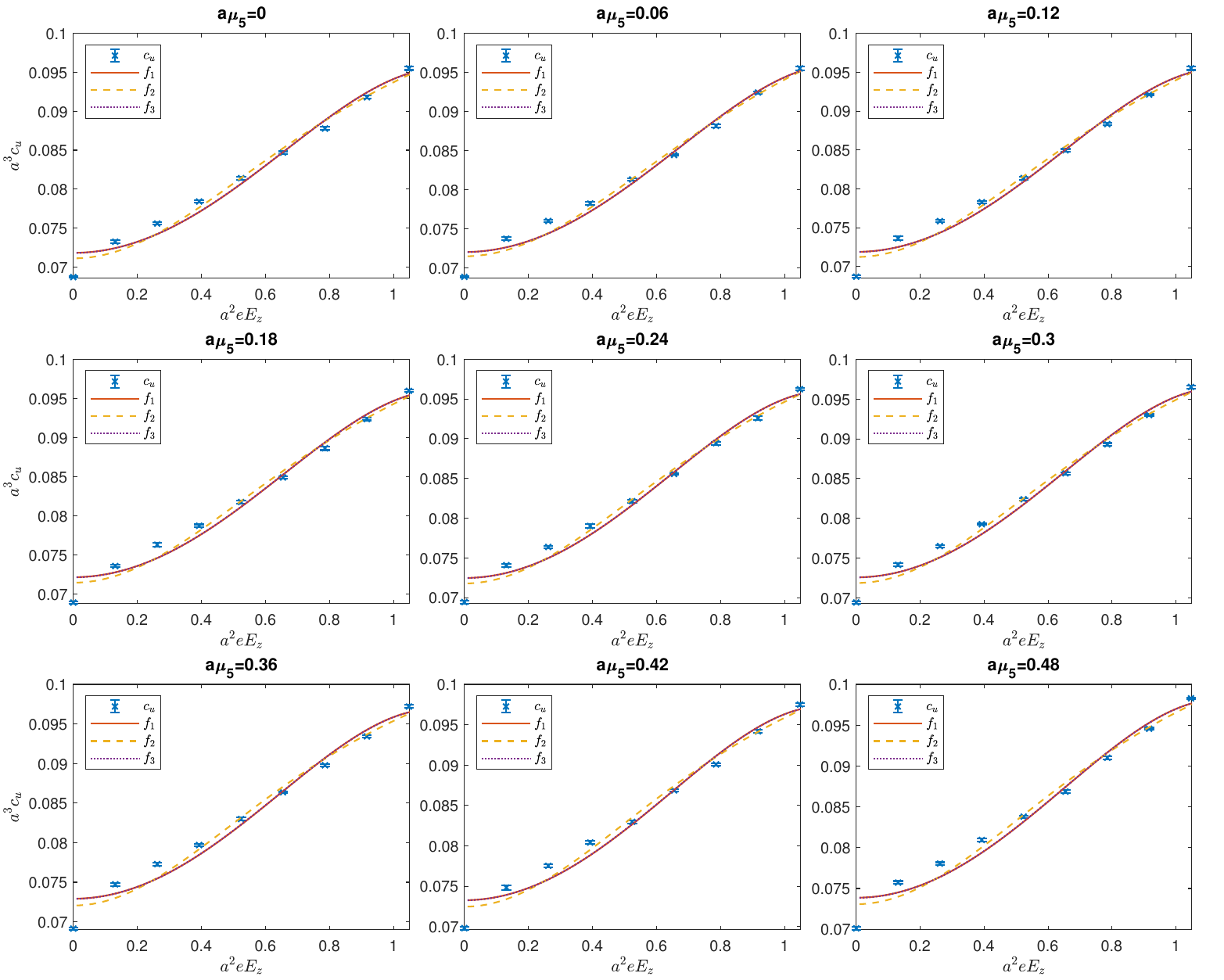}
\caption{\label{fig:chiralfit540u}
Same as Fig.~\ref{fig:chiralfit530u} but for $\beta=5.4, T=270.5\;{\rm MeV}$.}
\end{center}
\end{figure*}
\begin{figure*}[htbp]
\begin{center}
\includegraphics[width=0.98\hsize]{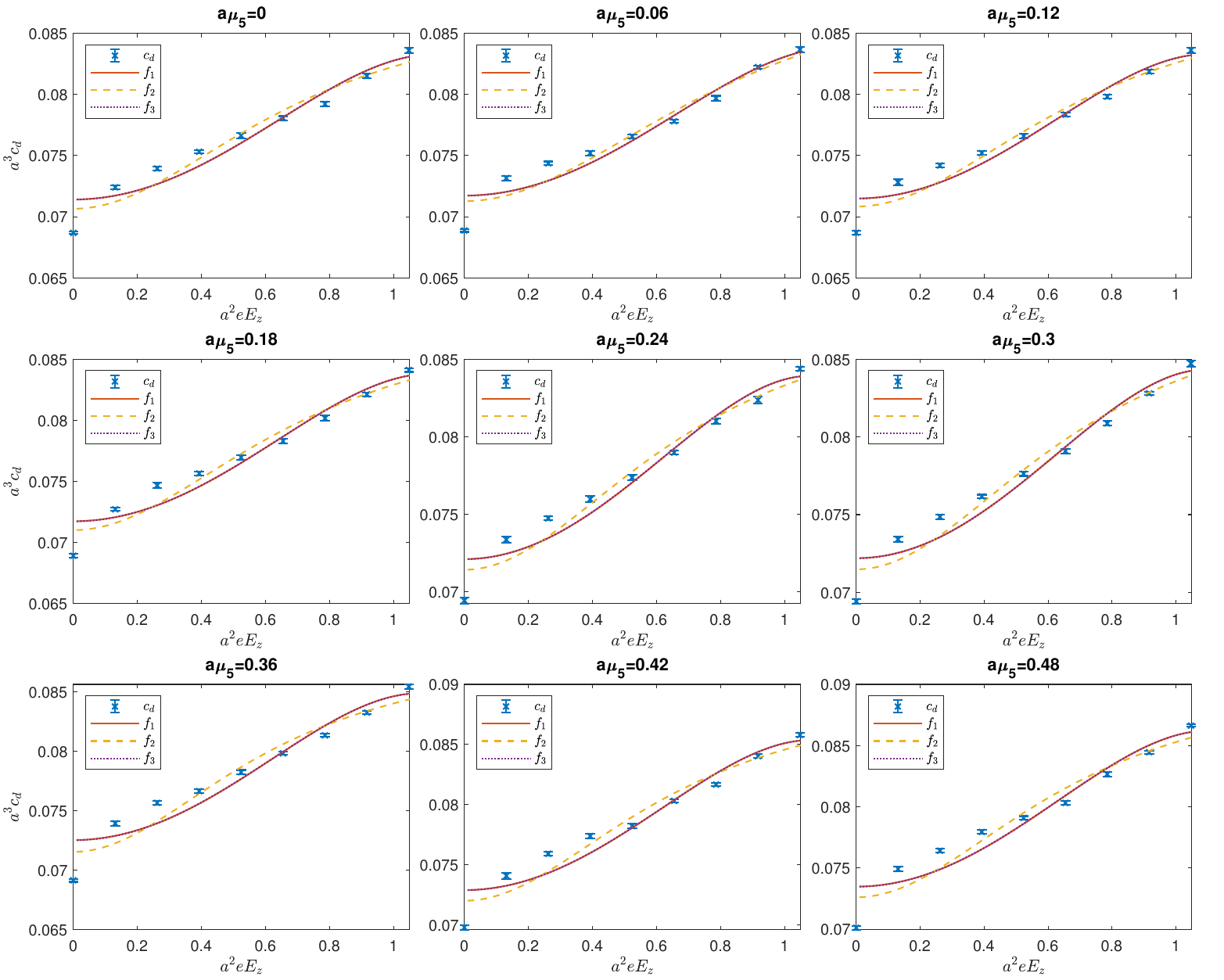}
\caption{\label{fig:chiralfit540d}
Same as Fig.~\ref{fig:chiralfit540u} but for $c_d$.}
\end{center}
\end{figure*}
Before Wick rotate the imaginary electric field back to real electric field, the chiral condensations as functions of $a^2eE_z$ and $\mu _5$ are fitted according to three different ansatzes listed in Eq.~(\ref{eq.3.chiralansatz}).
It can be found that for large $a^2eE_z$, the ansatzes fit well.
However, for small $a^2eE_z$, there are significant discrepancies due to the non-monotonic behavior, which indicate that new ansatzes based on theoretical analysis are need in the case that both the external electric field and the chiral chemical potential are presented.

\section{\label{sec:ap2}Fit of the Polyakov loop as functions of z-coordinate.}

\begin{figure*}[htbp]
\begin{center}
\includegraphics[width=0.98\hsize]{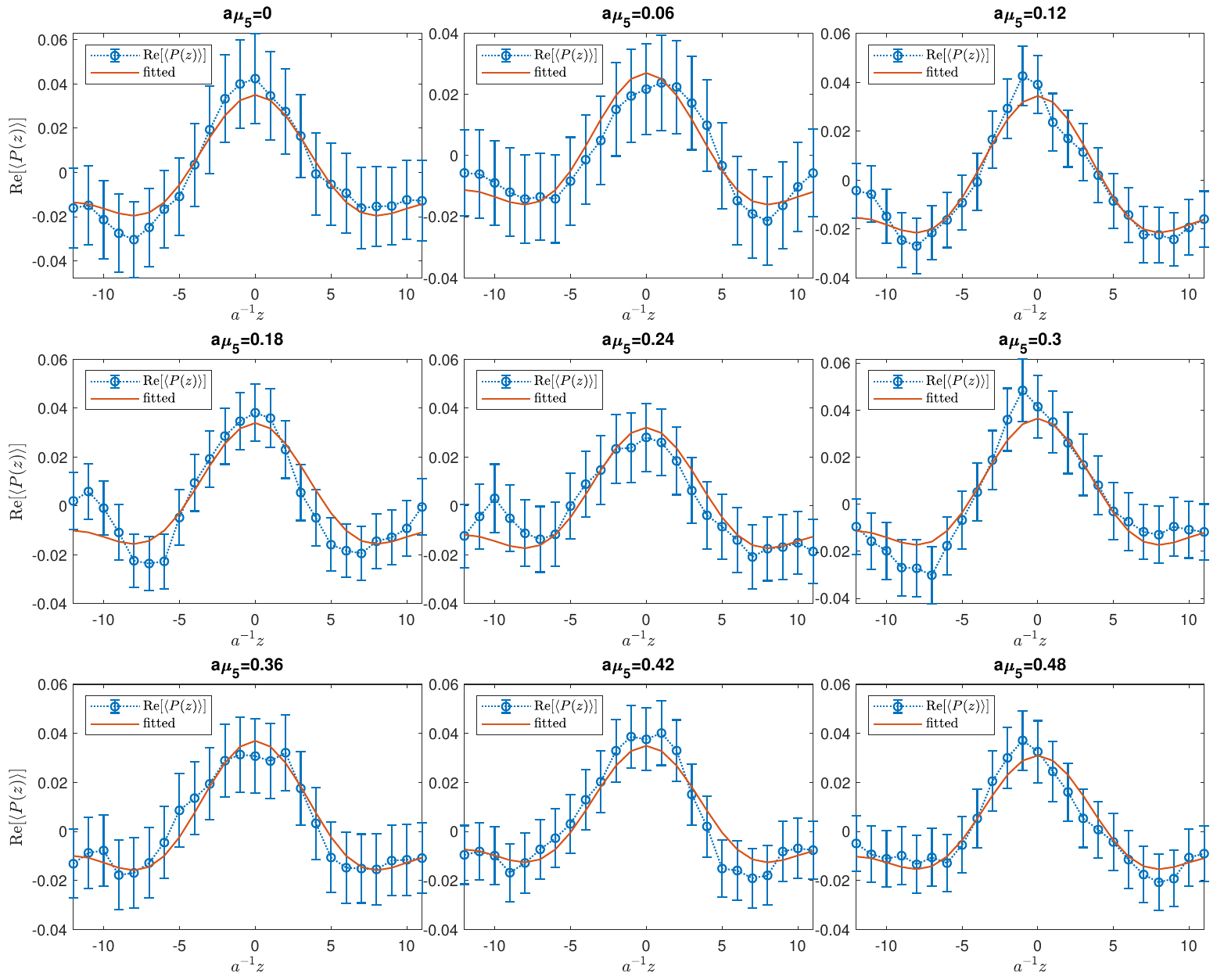}
\caption{\label{fig:530e1pzre}
${\rm Re}[\langle P(z) \rangle]$ as a function of $z$ coordinate at $\beta=5.3, T=202.5\;{\rm MeV}$ and $a^2eE_z=\pi/24$ with different chiral chemical potentials.
The result fitted with anstaz in Eq.~(\ref{eq.3.4}) are shown as the solid lines.}
\end{center}
\end{figure*}
\begin{figure*}[htbp]
\begin{center}
\includegraphics[width=0.98\hsize]{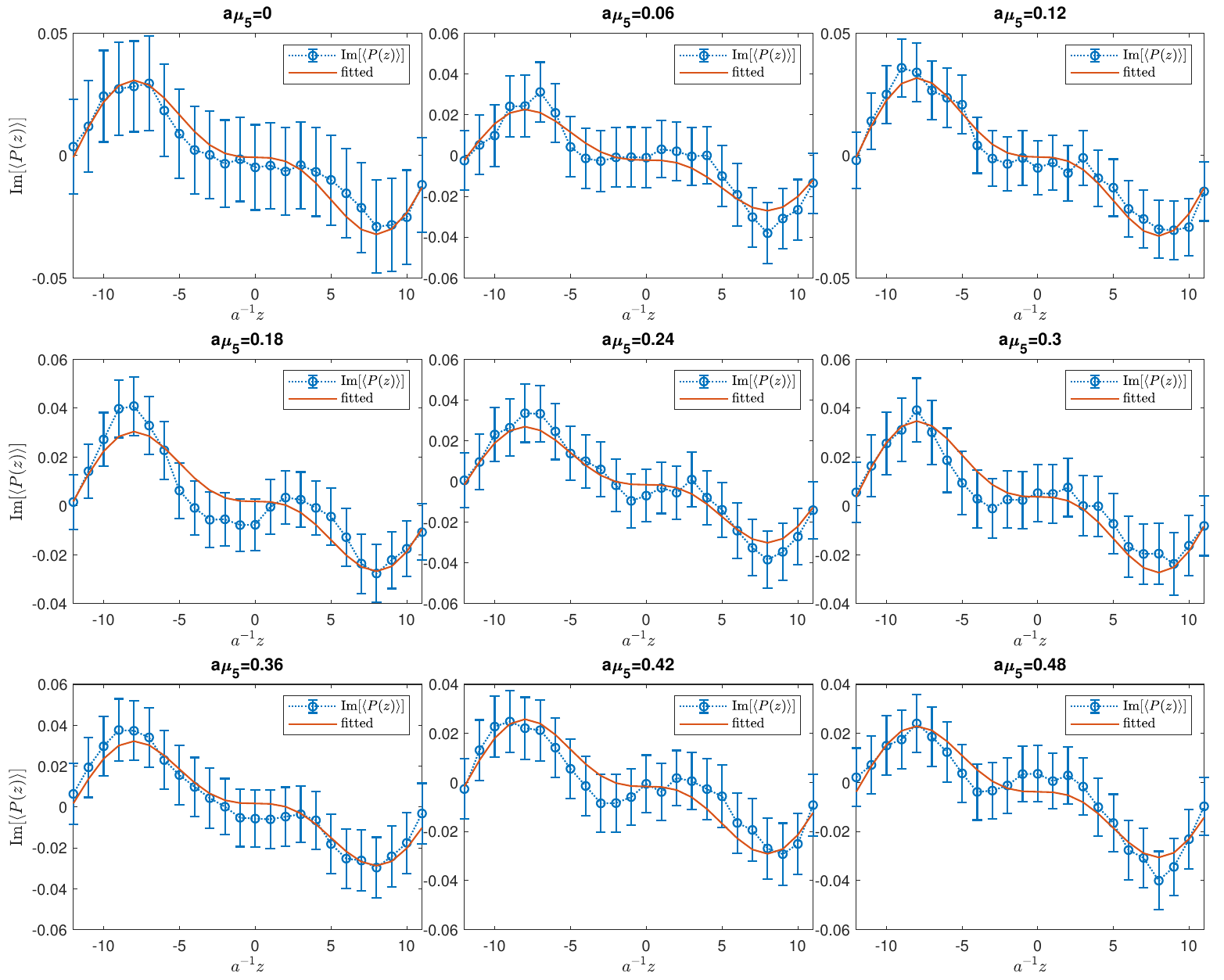}
\caption{\label{fig:530e1pzim}
Same as Fig.~\ref{fig:530e1pzre} but for ${\bf Im}[\langle P(z)\rangle]$.}
\end{center}
\end{figure*}
\begin{figure*}[htbp]
\begin{center}
\includegraphics[width=0.98\hsize]{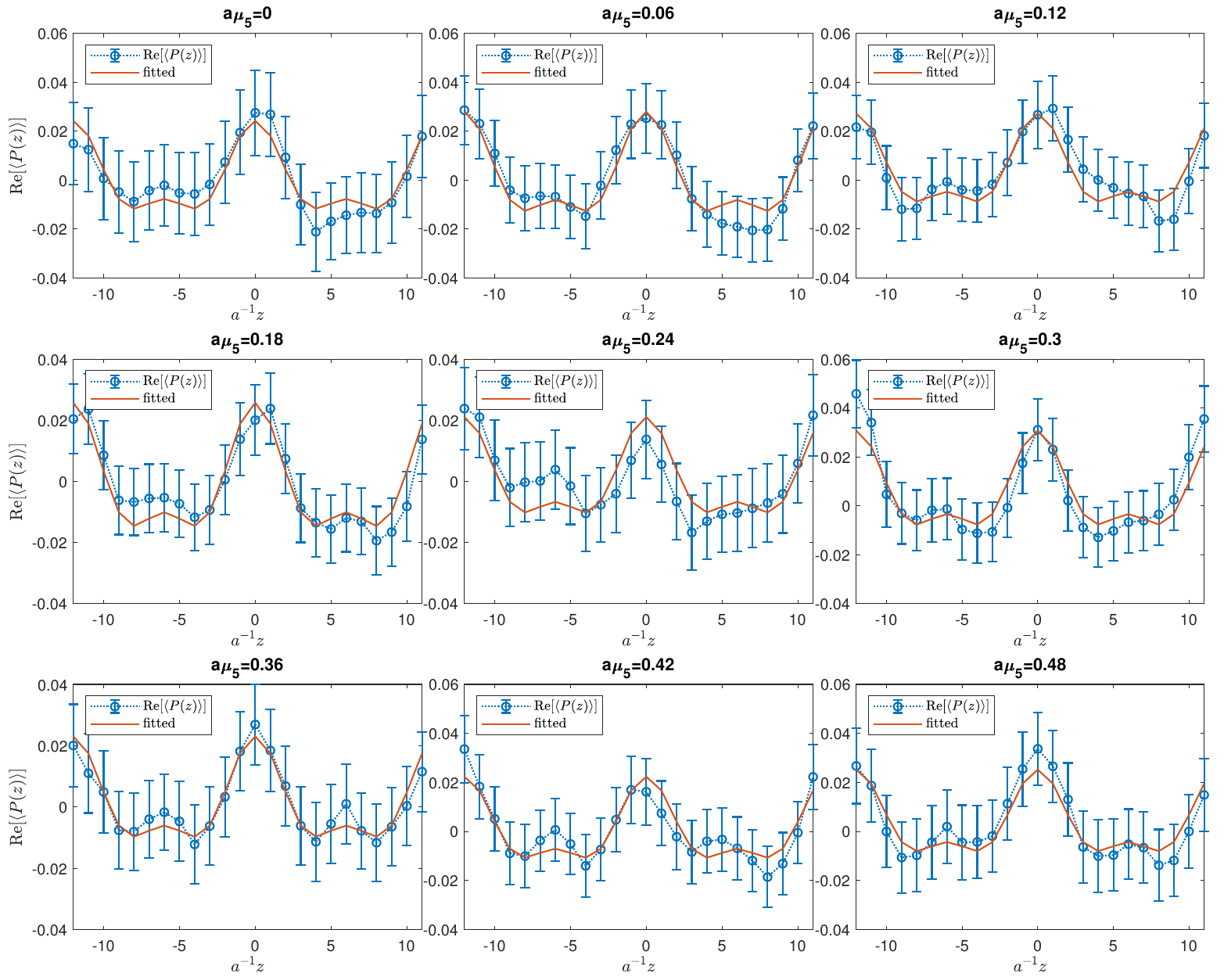}
\caption{\label{fig:530e2pzre}
Same as Fig.~\ref{fig:530e1pzre} but for $a^2eE_z=\pi/12$.}
\end{center}
\end{figure*}
\begin{figure*}[htbp]
\begin{center}
\includegraphics[width=0.98\hsize]{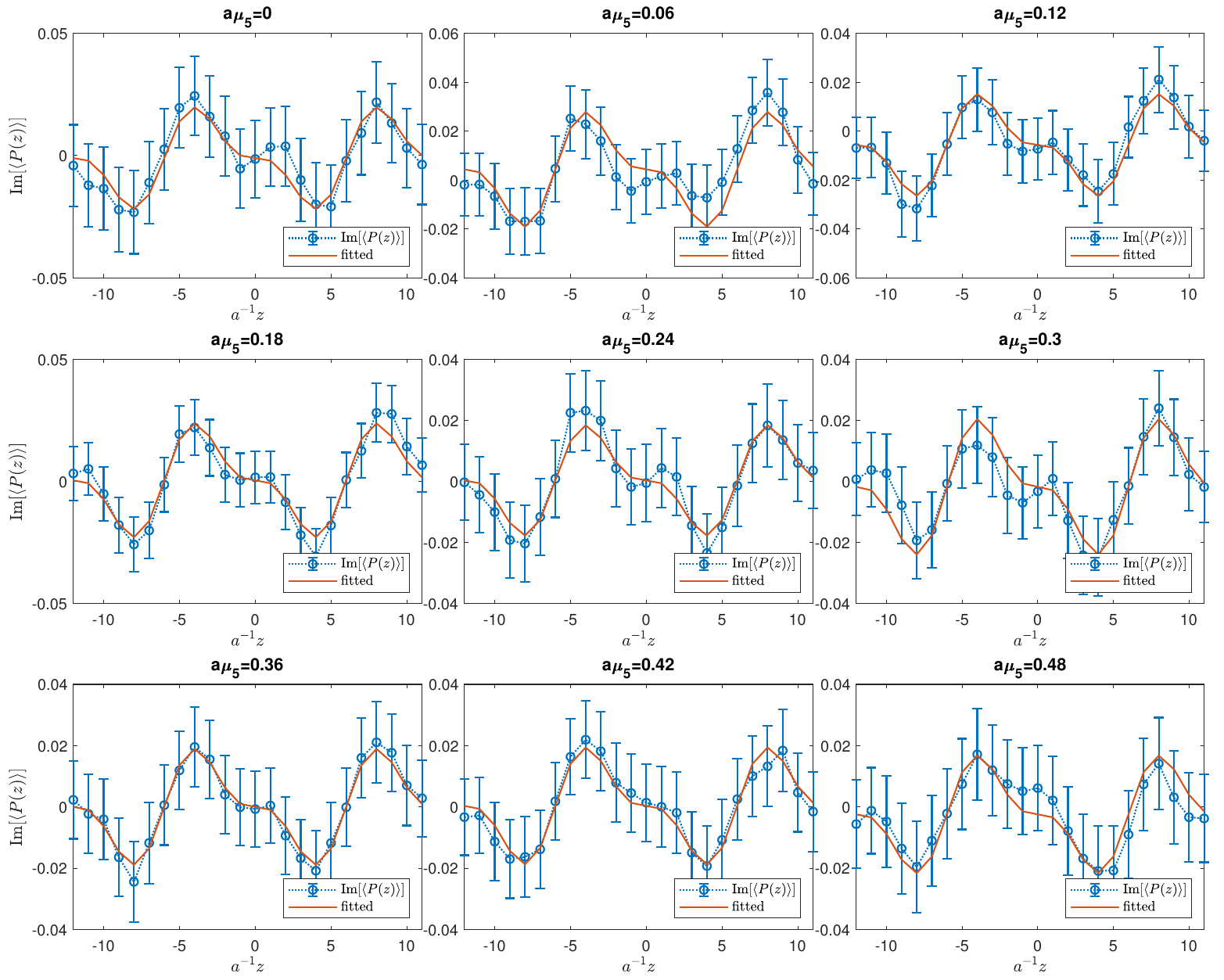}
\caption{\label{fig:530e2pzim}
Same as Fig.~\ref{fig:530e1pzim} but for $a^2eE_z=\pi/12$.}
\end{center}
\end{figure*}
\begin{figure*}[htbp]
\begin{center}
\includegraphics[width=0.98\hsize]{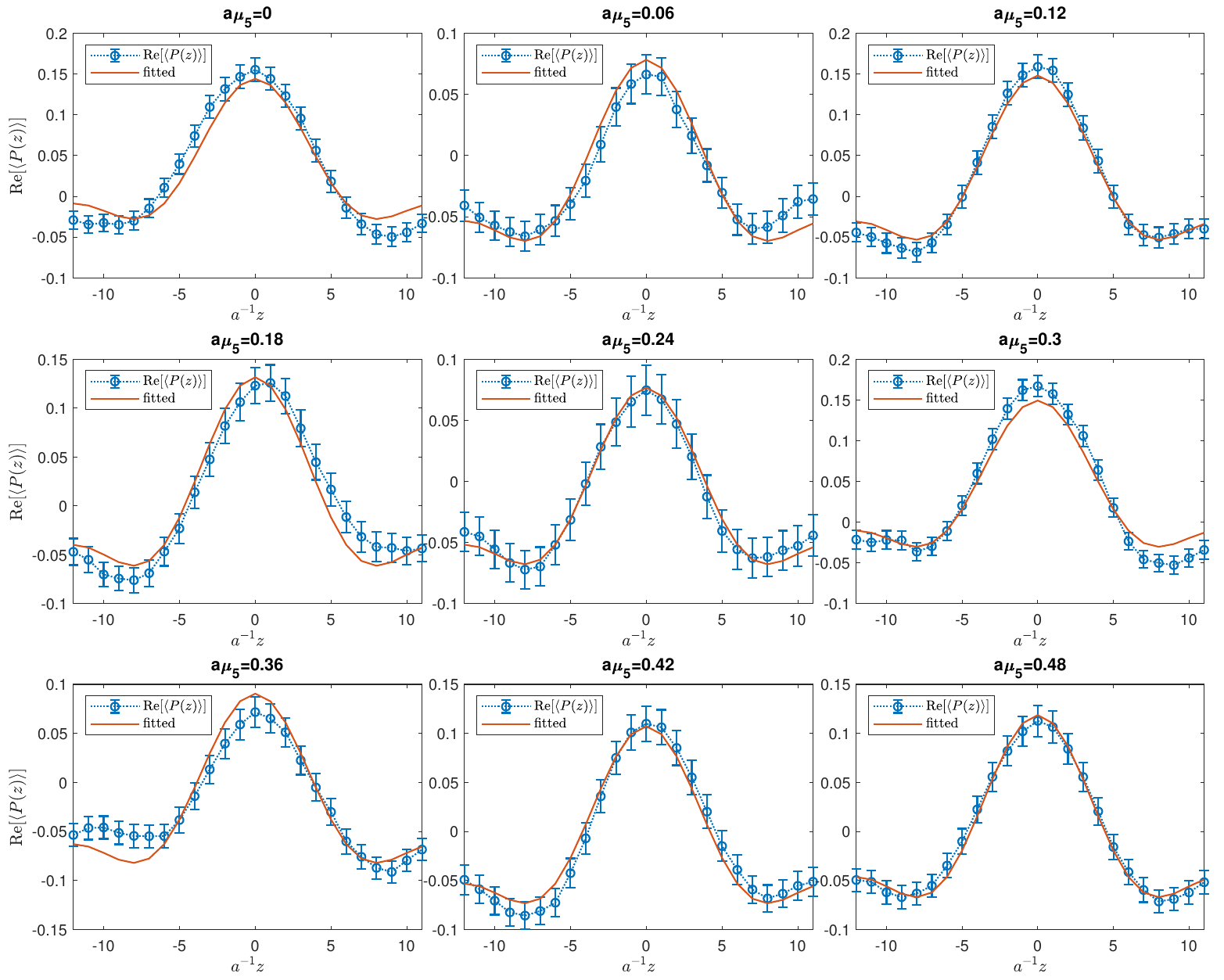}
\caption{\label{fig:540e1pzre}
Same as Fig.~\ref{fig:530e1pzre} but for $\beta=5.4, T=270.5\;{\rm MeV}$.}
\end{center}
\end{figure*}
\begin{figure*}[htbp]
\begin{center}
\includegraphics[width=0.98\hsize]{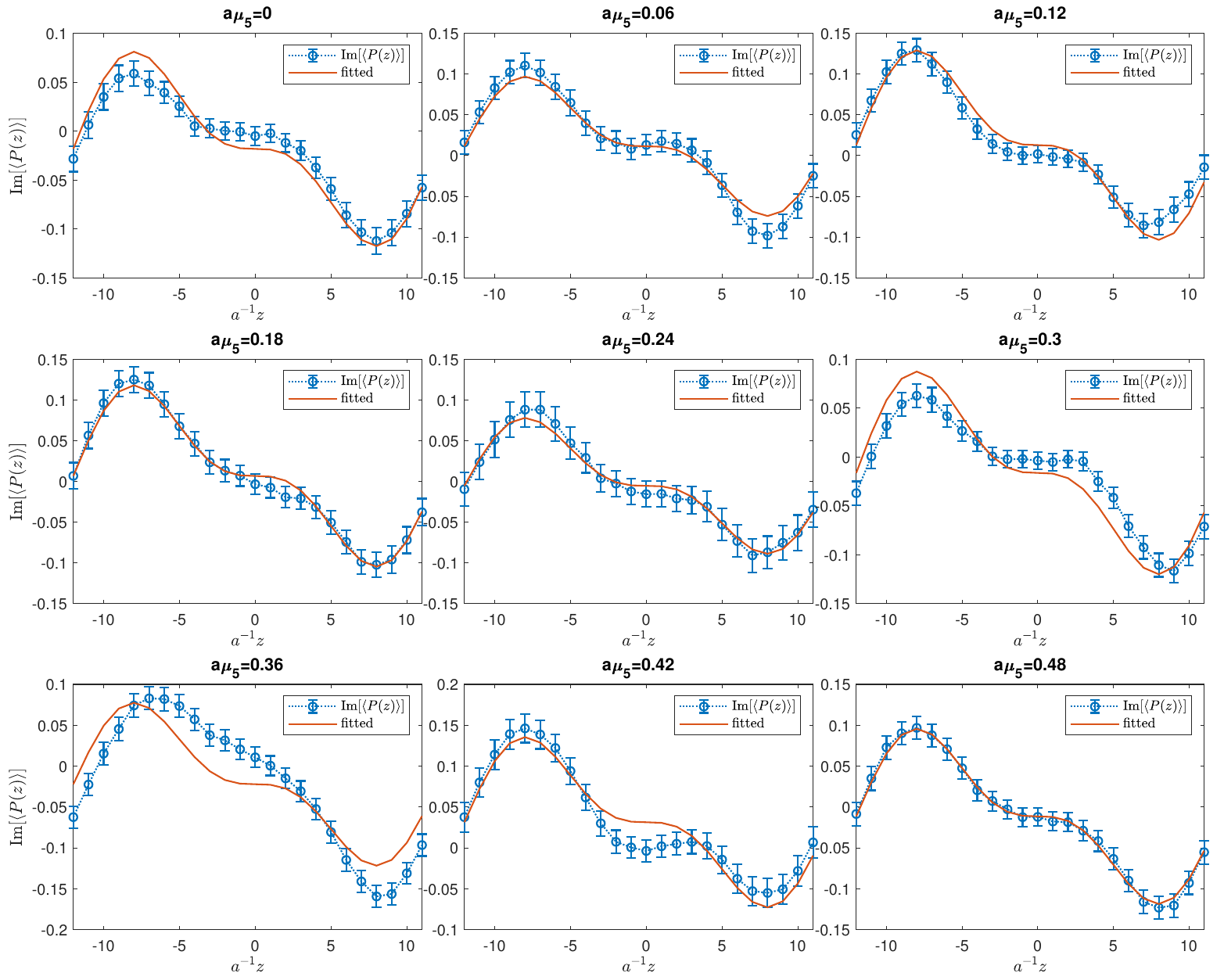}
\caption{\label{fig:540e1pzim}
Same as Fig.~\ref{fig:530e1pzim} but for $\beta=5.4, T=270.5\;{\rm MeV}$.}
\end{center}
\end{figure*}
\begin{figure*}[htbp]
\begin{center}
\includegraphics[width=0.98\hsize]{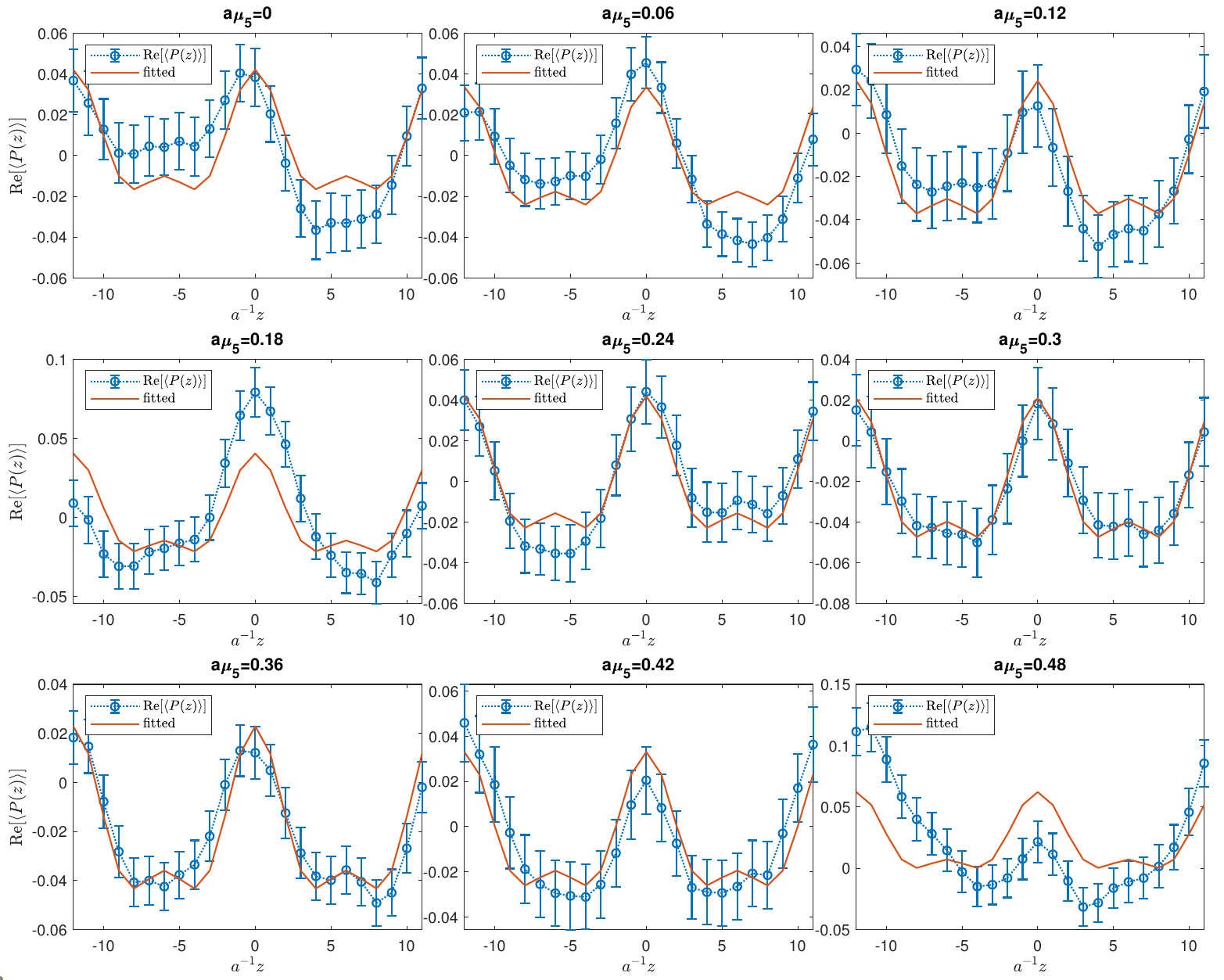}
\caption{\label{fig:540e2pzre}
Same as Fig.~\ref{fig:540e1pzre} but for $a^2eE_z=\pi/12$.}
\end{center}
\end{figure*}
\begin{figure*}[htbp]
\begin{center}
\includegraphics[width=0.98\hsize]{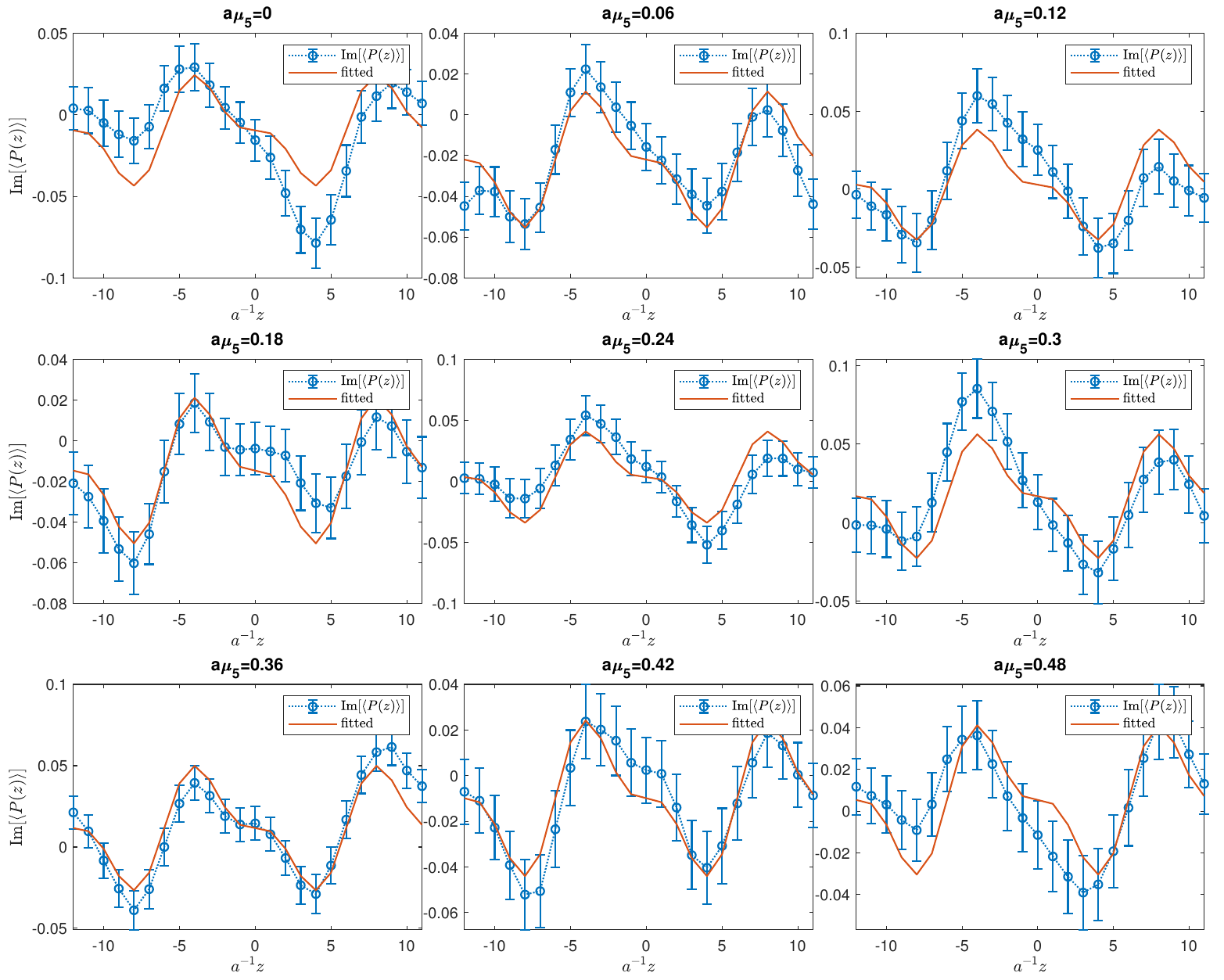}
\caption{\label{fig:540e2pzim}
Same as Fig.~\ref{fig:540e1pzim} but for $a^2eE_z=\pi/12$.}
\end{center}
\end{figure*}

Since the imaginary electric field acts like imaginary chemical potential varies along the ${\bf z}$-axis, there is a R-W transition.
It can shown that, in this case the Polyakov loop can be fitted using Eq.~(\ref{eq.3.4}).
The results for $\beta=5.3$ and $5.4$, for $a^2eE_z=\pi/24$ and $a^2eE_z=\pi/12$ are shown Figs.~\ref{fig:530e1pzre}-\ref{fig:540e2pzim}.
It can be seen that the Polyakov loop fits the anstaz well no matter whether the chiral chemical potential presents.

\bibliography{chiralpotential}

\begin{thebibliography}{10}
\expandafter\ifx\csname url\endcsname\relax
  \def\url#1{\texttt{#1}}\fi
\expandafter\ifx\csname urlprefix\endcsname\relax\def\urlprefix{URL }\fi
\expandafter\ifx\csname href\endcsname\relax
  \def\href#1#2{#2} \def\path#1{#1}\fi

\bibitem{Adler:1969gk}
S.~L. Adler, {Axial vector vertex in spinor electrodynamics}, Phys. Rev. 177
  (1969) 2426--2438.
\newblock \href {https://doi.org/10.1103/PhysRev.177.2426}
  {\path{doi:10.1103/PhysRev.177.2426}}.

\bibitem{Bell:1969ts}
J.~S. Bell, R.~Jackiw, {A PCAC puzzle: $\pi^0 \to \gamma \gamma$ in the
  $\sigma$ model}, Nuovo Cim. A 60 (1969) 47--61.
\newblock \href {https://doi.org/10.1007/BF02823296}
  {\path{doi:10.1007/BF02823296}}.

\bibitem{Andrianov:2016qgy}
A.~Andrianov, V.~Andrianov, D.~Espriu, {Chiral Imbalance in QCD and its
  consequences}, EPJ Web Conf. 125 (2016) 01009.
\newblock \href {https://doi.org/10.1051/epjconf/201612501009}
  {\path{doi:10.1051/epjconf/201612501009}}.

\bibitem{Klinkhamer:1984di}
F.~R. Klinkhamer, N.~S. Manton, {A Saddle Point Solution in the Weinberg-Salam
  Theory}, Phys. Rev. D 30 (1984) 2212.
\newblock \href {https://doi.org/10.1103/PhysRevD.30.2212}
  {\path{doi:10.1103/PhysRevD.30.2212}}.

\bibitem{Kuzmin:1985mm}
V.~A. Kuzmin, V.~A. Rubakov, M.~E. Shaposhnikov, {On the Anomalous Electroweak
  Baryon Number Nonconservation in the Early Universe}, Phys. Lett. B 155
  (1985) 36.
\newblock \href {https://doi.org/10.1016/0370-2693(85)91028-7}
  {\path{doi:10.1016/0370-2693(85)91028-7}}.

\bibitem{McLerran:1990de}
L.~D. McLerran, E.~Mottola, M.~E. Shaposhnikov, {Sphalerons and Axion Dynamics
  in High Temperature {QCD}}, Phys. Rev. D 43 (1991) 2027--2035.
\newblock \href {https://doi.org/10.1103/PhysRevD.43.2027}
  {\path{doi:10.1103/PhysRevD.43.2027}}.

\bibitem{Moore:1999fs}
G.~D. Moore, K.~Rummukainen, {Classical sphaleron rate on fine lattices}, Phys.
  Rev. D 61 (2000) 105008.
\newblock \href {http://arxiv.org/abs/hep-ph/9906259}
  {\path{arXiv:hep-ph/9906259}}, \href
  {https://doi.org/10.1103/PhysRevD.61.105008}
  {\path{doi:10.1103/PhysRevD.61.105008}}.

\bibitem{Shuryak:2002qz}
E.~Shuryak, I.~Zahed, {Prompt quark production by exploding sphalerons}, Phys.
  Rev. D 67 (2003) 014006.
\newblock \href {http://arxiv.org/abs/hep-ph/0206022}
  {\path{arXiv:hep-ph/0206022}}, \href
  {https://doi.org/10.1103/PhysRevD.67.014006}
  {\path{doi:10.1103/PhysRevD.67.014006}}.

\bibitem{Fukushima:2008xe}
K.~Fukushima, D.~E. Kharzeev, H.~J. Warringa, {The Chiral Magnetic Effect},
  Phys. Rev. D 78 (2008) 074033.
\newblock \href {http://arxiv.org/abs/0808.3382} {\path{arXiv:0808.3382}},
  \href {https://doi.org/10.1103/PhysRevD.78.074033}
  {\path{doi:10.1103/PhysRevD.78.074033}}.

\bibitem{Chernodub:2011fr}
M.~N. Chernodub, A.~S. Nedelin, {Phase diagram of chirally imbalanced QCD
  matter}, Phys. Rev. D 83 (2011) 105008.
\newblock \href {http://arxiv.org/abs/1102.0188} {\path{arXiv:1102.0188}},
  \href {https://doi.org/10.1103/PhysRevD.83.105008}
  {\path{doi:10.1103/PhysRevD.83.105008}}.

\bibitem{Kharzeev:2004ey}
D.~Kharzeev, {Parity violation in hot QCD: Why it can happen, and how to look
  for it}, Phys. Lett. B 633 (2006) 260--264.
\newblock \href {http://arxiv.org/abs/hep-ph/0406125}
  {\path{arXiv:hep-ph/0406125}}, \href
  {https://doi.org/10.1016/j.physletb.2005.11.075}
  {\path{doi:10.1016/j.physletb.2005.11.075}}.

\bibitem{Yu:2015hym}
L.~Yu, H.~Liu, M.~Huang, {Effect of the chiral chemical potential on the chiral
  phase transition in the NJL model with different regularization schemes},
  Phys. Rev. D 94~(1) (2016) 014026.
\newblock \href {http://arxiv.org/abs/1511.03073} {\path{arXiv:1511.03073}},
  \href {https://doi.org/10.1103/PhysRevD.94.014026}
  {\path{doi:10.1103/PhysRevD.94.014026}}.

\bibitem{Fukushima:2010fe}
K.~Fukushima, M.~Ruggieri, R.~Gatto, {Chiral magnetic effect in the PNJL
  model}, Phys. Rev. D 81 (2010) 114031.
\newblock \href {http://arxiv.org/abs/1003.0047} {\path{arXiv:1003.0047}},
  \href {https://doi.org/10.1103/PhysRevD.81.114031}
  {\path{doi:10.1103/PhysRevD.81.114031}}.

\bibitem{Gatto:2011wc}
R.~Gatto, M.~Ruggieri, {Hot Quark Matter with an Axial Chemical Potential},
  Phys. Rev. D 85 (2012) 054013.
\newblock \href {http://arxiv.org/abs/1110.4904} {\path{arXiv:1110.4904}},
  \href {https://doi.org/10.1103/PhysRevD.85.054013}
  {\path{doi:10.1103/PhysRevD.85.054013}}.

\bibitem{Ruggieri:2011xc}
M.~Ruggieri, {The Critical End Point of Quantum Chromodynamics Detected by
  Chirally Imbalanced Quark Matter}, Phys. Rev. D 84 (2011) 014011.
\newblock \href {http://arxiv.org/abs/1103.6186} {\path{arXiv:1103.6186}},
  \href {https://doi.org/10.1103/PhysRevD.84.014011}
  {\path{doi:10.1103/PhysRevD.84.014011}}.

\bibitem{Espriu:2020dge}
D.~Espriu, A.~G\'omez~Nicola, A.~Vioque-Rodr\'\i{}guez, {Chiral perturbation
  theory for nonzero chiral imbalance}, JHEP 06 (2020) 062.
\newblock \href {http://arxiv.org/abs/2002.11696} {\path{arXiv:2002.11696}},
  \href {https://doi.org/10.1007/JHEP06(2020)062}
  {\path{doi:10.1007/JHEP06(2020)062}}.

\bibitem{Shi:2020uyb}
C.~Shi, X.-T. He, W.-B. Jia, Q.-W. Wang, S.-S. Xu, H.-S. Zong, {Chiral
  transition and the chiral charge density of the hot and dense QCD matter},
  JHEP 06 (2020) 122.
\newblock \href {http://arxiv.org/abs/2004.09918} {\path{arXiv:2004.09918}},
  \href {https://doi.org/10.1007/JHEP06(2020)122}
  {\path{doi:10.1007/JHEP06(2020)122}}.

\bibitem{Bzdak:2011yy}
A.~Bzdak, V.~Skokov, {Event-by-event fluctuations of magnetic and electric
  fields in heavy ion collisions}, Phys. Lett. B 710 (2012) 171--174.
\newblock \href {http://arxiv.org/abs/1111.1949} {\path{arXiv:1111.1949}},
  \href {https://doi.org/10.1016/j.physletb.2012.02.065}
  {\path{doi:10.1016/j.physletb.2012.02.065}}.

\bibitem{Deng:2012pc}
W.-T. Deng, X.-G. Huang, {Event-by-event generation of electromagnetic fields
  in heavy-ion collisions}, Phys. Rev. C 85 (2012) 044907.
\newblock \href {http://arxiv.org/abs/1201.5108} {\path{arXiv:1201.5108}},
  \href {https://doi.org/10.1103/PhysRevC.85.044907}
  {\path{doi:10.1103/PhysRevC.85.044907}}.

\bibitem{Bloczynski:2012en}
J.~Bloczynski, X.-G. Huang, X.~Zhang, J.~Liao, {Azimuthally fluctuating
  magnetic field and its impacts on observables in heavy-ion collisions}, Phys.
  Lett. B 718 (2013) 1529--1535.
\newblock \href {http://arxiv.org/abs/1209.6594} {\path{arXiv:1209.6594}},
  \href {https://doi.org/10.1016/j.physletb.2012.12.030}
  {\path{doi:10.1016/j.physletb.2012.12.030}}.

\bibitem{Hirono:2012rt}
Y.~Hirono, M.~Hongo, T.~Hirano, {Estimation of electric conductivity of the
  quark gluon plasma via asymmetric heavy-ion collisions}, Phys. Rev. C 90~(2)
  (2014) 021903.
\newblock \href {http://arxiv.org/abs/1211.1114} {\path{arXiv:1211.1114}},
  \href {https://doi.org/10.1103/PhysRevC.90.021903}
  {\path{doi:10.1103/PhysRevC.90.021903}}.

\bibitem{Deng:2014uja}
W.-T. Deng, X.-G. Huang, {Electric fields and chiral magnetic effect in Cu+Au
  collisions}, Phys. Lett. B 742 (2015) 296--302.
\newblock \href {http://arxiv.org/abs/1411.2733} {\path{arXiv:1411.2733}},
  \href {https://doi.org/10.1016/j.physletb.2015.01.050}
  {\path{doi:10.1016/j.physletb.2015.01.050}}.

\bibitem{Voronyuk:2014rna}
V.~Voronyuk, V.~D. Toneev, S.~A. Voloshin, W.~Cassing, {Charge-dependent
  directed flow in asymmetric nuclear collisions}, Phys. Rev. C 90~(6) (2014)
  064903.
\newblock \href {http://arxiv.org/abs/1410.1402} {\path{arXiv:1410.1402}},
  \href {https://doi.org/10.1103/PhysRevC.90.064903}
  {\path{doi:10.1103/PhysRevC.90.064903}}.

\bibitem{Babansky:1997zh}
A.~Y. Babansky, E.~V. Gorbar, G.~V. Shchepanyuk, {Chiral symmetry breaking in
  the Nambu-Jona-Lasinio model in external constant electromagnetic field},
  Phys. Lett. B 419 (1998) 272--278.
\newblock \href {http://arxiv.org/abs/hep-th/9705218}
  {\path{arXiv:hep-th/9705218}}, \href
  {https://doi.org/10.1016/S0370-2693(97)01445-7}
  {\path{doi:10.1016/S0370-2693(97)01445-7}}.

\bibitem{Klevansky:1989vi}
S.~P. Klevansky, R.~H. Lemmer, {Chiral symmetry restoration in the
  Nambu-Jona-Lasinio model with a constant electromagnetic field}, Phys. Rev. D
  39 (1989) 3478--3489.
\newblock \href {https://doi.org/10.1103/PhysRevD.39.3478}
  {\path{doi:10.1103/PhysRevD.39.3478}}.

\bibitem{Suganuma:1990nn}
H.~Suganuma, T.~Tatsumi, {On the Behavior of Symmetry and Phase Transitions in
  a Strong Electromagnetic Field}, Annals Phys. 208 (1991) 470--508.
\newblock \href {https://doi.org/10.1016/0003-4916(91)90304-Q}
  {\path{doi:10.1016/0003-4916(91)90304-Q}}.

\bibitem{Tavares:2019mvq}
W.~R. Tavares, R.~L.~S. Farias, S.~S. Avancini, {Deconfinement and chiral phase
  transitions in quark matter with a strong electric field}, Phys. Rev. D
  101~(1) (2020) 016017.
\newblock \href {http://arxiv.org/abs/1912.00305} {\path{arXiv:1912.00305}},
  \href {https://doi.org/10.1103/PhysRevD.101.016017}
  {\path{doi:10.1103/PhysRevD.101.016017}}.

\bibitem{Cao:2015dya}
G.~Cao, X.-G. Huang, {Chiral phase transition and Schwinger mechanism in a pure
  electric field}, Phys. Rev. D 93~(1) (2016) 016007.
\newblock \href {http://arxiv.org/abs/1510.05125} {\path{arXiv:1510.05125}},
  \href {https://doi.org/10.1103/PhysRevD.93.016007}
  {\path{doi:10.1103/PhysRevD.93.016007}}.

\bibitem{Ruggieri:2016xww}
M.~Ruggieri, Z.~Y. Lu, G.~X. Peng, {Influence of chiral chemical potential,
  parallel electric, and magnetic fields on the critical temperature of QCD},
  Phys. Rev. D 94~(11) (2016) 116003.
\newblock \href {http://arxiv.org/abs/1608.08310} {\path{arXiv:1608.08310}},
  \href {https://doi.org/10.1103/PhysRevD.94.116003}
  {\path{doi:10.1103/PhysRevD.94.116003}}.

\bibitem{Ruggieri:2016jrt}
M.~Ruggieri, G.-X. Peng, {Chiral phase transition of quark matter in the
  background of parallel electric and magnetic fields}, Nucl. Sci. Tech. 27~(6)
  (2016) 130.
\newblock \href {https://doi.org/10.1007/s41365-016-0139-x}
  {\path{doi:10.1007/s41365-016-0139-x}}.

\bibitem{Ruggieri:2016lrn}
M.~Ruggieri, G.~X. Peng, {Quark matter in a parallel electric and magnetic
  field background: Chiral phase transition and equilibration of chiral
  density}, Phys. Rev. D 93~(9) (2016) 094021.
\newblock \href {http://arxiv.org/abs/1602.08994} {\path{arXiv:1602.08994}},
  \href {https://doi.org/10.1103/PhysRevD.93.094021}
  {\path{doi:10.1103/PhysRevD.93.094021}}.

\bibitem{Brandt:2022jfk}
B.~Brandt, F.~Cuteri, G.~Endr\H{o}di, J.~J.~H. Hern\'andez, G.~Mark\'o, {QCD
  topology with electromagnetic fields and the axion-photon coupling}, PoS
  LATTICE2022 (2023) 174.
\newblock \href {http://arxiv.org/abs/2212.03385} {\path{arXiv:2212.03385}},
  \href {https://doi.org/10.22323/1.430.0174} {\path{doi:10.22323/1.430.0174}}.

\bibitem{Yamamoto:2011gk}
A.~Yamamoto, {Chiral magnetic effect in lattice QCD with a chiral chemical
  potential}, Phys. Rev. Lett. 107 (2011) 031601.
\newblock \href {http://arxiv.org/abs/1105.0385} {\path{arXiv:1105.0385}},
  \href {https://doi.org/10.1103/PhysRevLett.107.031601}
  {\path{doi:10.1103/PhysRevLett.107.031601}}.

\bibitem{Yamamoto:2011ks}
A.~Yamamoto, {Lattice study of the chiral magnetic effect in a chirally
  imbalanced matter}, Phys. Rev. D 84 (2011) 114504.
\newblock \href {http://arxiv.org/abs/1111.4681} {\path{arXiv:1111.4681}},
  \href {https://doi.org/10.1103/PhysRevD.84.114504}
  {\path{doi:10.1103/PhysRevD.84.114504}}.

\bibitem{Braguta:2014ira}
V.~V. Braguta, E.~M. Ilgenfritz, A.~Y. Kotov, M.~M\"uller-Preussker,
  B.~Petersson, A.~Schreiber, {Two-Color QCD with Chiral Chemical Potential},
  PoS LATTICE2014 (2015) 235.
\newblock \href {http://arxiv.org/abs/1411.5174} {\path{arXiv:1411.5174}},
  \href {https://doi.org/10.22323/1.214.0235} {\path{doi:10.22323/1.214.0235}}.

\bibitem{Braguta:2015zta}
V.~V. Braguta, V.~A. Goy, E.~M. Ilgenfritz, A.~Y. Kotov, A.~V. Molochkov,
  M.~Muller-Preussker, B.~Petersson, {Two-Color QCD with Non-zero Chiral
  Chemical Potential}, JHEP 06 (2015) 094.
\newblock \href {http://arxiv.org/abs/1503.06670} {\path{arXiv:1503.06670}},
  \href {https://doi.org/10.1007/JHEP06(2015)094}
  {\path{doi:10.1007/JHEP06(2015)094}}.

\bibitem{Kotov:2015hxr}
A.~Y. Kotov, V.~V. Braguta, V.~A. Goy, E.-M. Ilgenfritz, A.~Molochkov,
  M.~Muller-Preussker, B.~Petersson, S.~A. Skinderev, {Lattice QCD with Chiral
  Chemical Potential: from SU(2) to SU(3)}, PoS LATTICE2015 (2016) 185.
\newblock \href {https://doi.org/10.22323/1.251.0185}
  {\path{doi:10.22323/1.251.0185}}.

\bibitem{Braguta:2019pxt}
V.~V. Braguta, M.~I. Katsnelson, A.~Y. Kotov, A.~M. Trunin, {Catalysis of
  Dynamical Chiral Symmetry Breaking by Chiral Chemical Potential in Dirac
  semimetals}, Phys. Rev. B 100~(8) (2019) 085117.
\newblock \href {http://arxiv.org/abs/1904.07003} {\path{arXiv:1904.07003}},
  \href {https://doi.org/10.1103/PhysRevB.100.085117}
  {\path{doi:10.1103/PhysRevB.100.085117}}.

\bibitem{Astrakhantsev:2019wnp}
N.~Y. Astrakhantsev, V.~V. Braguta, A.~Y. Kotov, D.~D. Kuznedelev, A.~A.
  Nikolaev, {Lattice study of QCD at finite chiral density: topology and
  confinement}, Eur. Phys. J. A 57~(1) (2021) 15.
\newblock \href {http://arxiv.org/abs/1902.09325} {\path{arXiv:1902.09325}},
  \href {https://doi.org/10.1140/epja/s10050-020-00326-2}
  {\path{doi:10.1140/epja/s10050-020-00326-2}}.

\bibitem{Yamamoto:2012bd}
A.~Yamamoto, {Lattice QCD with strong external electric fields}, Phys. Rev.
  Lett. 110~(11) (2013) 112001.
\newblock \href {http://arxiv.org/abs/1210.8250} {\path{arXiv:1210.8250}},
  \href {https://doi.org/10.1103/PhysRevLett.110.112001}
  {\path{doi:10.1103/PhysRevLett.110.112001}}.

\bibitem{Shintani:2006xr}
E.~Shintani, S.~Aoki, N.~Ishizuka, K.~Kanaya, Y.~Kikukawa, Y.~Kuramashi,
  M.~Okawa, A.~Ukawa, T.~Yoshie, {Neutron electric dipole moment with external
  electric field method in lattice QCD}, Phys. Rev. D 75 (2007) 034507.
\newblock \href {http://arxiv.org/abs/hep-lat/0611032}
  {\path{arXiv:hep-lat/0611032}}, \href
  {https://doi.org/10.1103/PhysRevD.75.034507}
  {\path{doi:10.1103/PhysRevD.75.034507}}.

\bibitem{Alexandru:2008sj}
A.~Alexandru, F.~X. Lee, {The Background field method on the lattice}, PoS
  LATTICE2008 (2008) 145.
\newblock \href {http://arxiv.org/abs/0810.2833} {\path{arXiv:0810.2833}},
  \href {https://doi.org/10.22323/1.066.0145} {\path{doi:10.22323/1.066.0145}}.

\bibitem{DElia:2012ifm}
M.~D'Elia, M.~Mariti, F.~Negro, {Susceptibility of the QCD vacuum to CP-odd
  electromagnetic background fields}, Phys. Rev. Lett. 110~(8) (2013) 082002.
\newblock \href {http://arxiv.org/abs/1209.0722} {\path{arXiv:1209.0722}},
  \href {https://doi.org/10.1103/PhysRevLett.110.082002}
  {\path{doi:10.1103/PhysRevLett.110.082002}}.

\bibitem{Fiebig:1988en}
H.~R. Fiebig, W.~Wilcox, R.~M. Woloshyn, {A Study of Hadron Electric
  Polarizability in Quenched Lattice {QCD}}, Nucl. Phys. B 324 (1989) 47--66.
\newblock \href {https://doi.org/10.1016/0550-3213(89)90180-6}
  {\path{doi:10.1016/0550-3213(89)90180-6}}.

\bibitem{Christensen:2004ca}
J.~C. Christensen, W.~Wilcox, F.~X. Lee, L.-m. Zhou, {Electric polarizability
  of neutral hadrons from lattice QCD}, Phys. Rev. D 72 (2005) 034503.
\newblock \href {http://arxiv.org/abs/hep-lat/0408024}
  {\path{arXiv:hep-lat/0408024}}, \href
  {https://doi.org/10.1103/PhysRevD.72.034503}
  {\path{doi:10.1103/PhysRevD.72.034503}}.

\bibitem{Engelhardt:2007ub}
M.~Engelhardt, {Neutron electric polarizability from unquenched lattice QCD
  using the background field approach}, Phys. Rev. D 76 (2007) 114502.
\newblock \href {http://arxiv.org/abs/0706.3919} {\path{arXiv:0706.3919}},
  \href {https://doi.org/10.1103/PhysRevD.76.114502}
  {\path{doi:10.1103/PhysRevD.76.114502}}.

\bibitem{Endrodi:2022wym}
G.~Endr\H{o}di, G.~Mark\'o, {On electric fields in hot QCD: perturbation
  theory}, JHEP 12 (2022) 015.
\newblock \href {http://arxiv.org/abs/2208.14306} {\path{arXiv:2208.14306}},
  \href {https://doi.org/10.1007/JHEP12(2022)015}
  {\path{doi:10.1007/JHEP12(2022)015}}.

\bibitem{Endrodi:2021qxz}
G.~Endrodi, G.~Marko, {Thermal QCD with external imaginary electric fields on
  the lattice}, PoS LATTICE2021 (2022) 245.
\newblock \href {http://arxiv.org/abs/2110.12189} {\path{arXiv:2110.12189}},
  \href {https://doi.org/10.22323/1.396.0245} {\path{doi:10.22323/1.396.0245}}.

\bibitem{Yang:2022zob}
J.-C. Yang, X.-T. Chang, J.-X. Chen, {Study of the Roberge-Weiss phase caused
  by external uniform classical electric field using lattice QCD approach},
  JHEP 10 (2022) 053.
\newblock \href {http://arxiv.org/abs/2207.11796} {\path{arXiv:2207.11796}},
  \href {https://doi.org/10.1007/JHEP10(2022)053}
  {\path{doi:10.1007/JHEP10(2022)053}}.

\bibitem{Roberge:1986mm}
A.~Roberge, N.~Weiss, {Gauge Theories With Imaginary Chemical Potential and the
  Phases of {QCD}}, Nucl. Phys. B 275 (1986) 734--745.
\newblock \href {https://doi.org/10.1016/0550-3213(86)90582-1}
  {\path{doi:10.1016/0550-3213(86)90582-1}}.

\bibitem{Philipsen:2014rpa}
O.~Philipsen, C.~Pinke, {Nature of the Roberge-Weiss transition in $N_f=2$ QCD
  with Wilson fermions}, Phys. Rev. D 89~(9) (2014) 094504.
\newblock \href {http://arxiv.org/abs/1402.0838} {\path{arXiv:1402.0838}},
  \href {https://doi.org/10.1103/PhysRevD.89.094504}
  {\path{doi:10.1103/PhysRevD.89.094504}}.

\bibitem{Wu:2013bfa}
L.-K. Wu, X.-F. Meng, {Nature of the Roberge-Weiss transition end points in
  two-flavor lattice QCD with Wilson quarks}, Phys. Rev. D 87~(9) (2013)
  094508.
\newblock \href {http://arxiv.org/abs/1303.0336} {\path{arXiv:1303.0336}},
  \href {https://doi.org/10.1103/PhysRevD.87.094508}
  {\path{doi:10.1103/PhysRevD.87.094508}}.

\bibitem{Wu:2014lsa}
L.-K. Wu, X.-F. Meng, {Nature of Roberge-Weiss transition endpoints for heavy
  quarks in $N_{f} =$ 2 lattice QCD with Wilson fermions}, Phys. Rev. D 90~(9)
  (2014) 094506.
\newblock \href {http://arxiv.org/abs/1405.2425} {\path{arXiv:1405.2425}},
  \href {https://doi.org/10.1103/PhysRevD.90.094506}
  {\path{doi:10.1103/PhysRevD.90.094506}}.

\bibitem{Cuteri:2015ayx}
F.~Cuteri, C.~Czaban, O.~Philipsen, C.~Pinke, A.~Sciarra, {The nature of the
  Roberge-Weiss Transition in $N_f=2$ QCD with Wilson Fermions on $N_t=6$
  lattices}, PoS LATTICE2015 (2016) 148.
\newblock \href {http://arxiv.org/abs/1511.03105} {\path{arXiv:1511.03105}},
  \href {https://doi.org/10.22323/1.251.0148} {\path{doi:10.22323/1.251.0148}}.

\bibitem{Bonati:2018fvg}
C.~Bonati, E.~Calore, M.~D'Elia, M.~Mesiti, F.~Negro, F.~Sanfilippo, S.~F.
  Schifano, G.~Silvi, R.~Tripiccione, {Roberge-Weiss endpoint and chiral
  symmetry restoration in $N_f = 2+1$ QCD}, Phys. Rev. D 99~(1) (2019) 014502.
\newblock \href {http://arxiv.org/abs/1807.02106} {\path{arXiv:1807.02106}},
  \href {https://doi.org/10.1103/PhysRevD.99.014502}
  {\path{doi:10.1103/PhysRevD.99.014502}}.

\bibitem{Cuteri:2022vwk}
F.~Cuteri, J.~Goswami, F.~Karsch, A.~Lahiri, M.~Neumann, O.~Philipsen,
  C.~Schmidt, A.~Sciarra, {Toward the chiral phase transition in the
  Roberge-Weiss plane}, Phys. Rev. D 106~(1) (2022) 014510.
\newblock \href {http://arxiv.org/abs/2205.12707} {\path{arXiv:2205.12707}},
  \href {https://doi.org/10.1103/PhysRevD.106.014510}
  {\path{doi:10.1103/PhysRevD.106.014510}}.

\bibitem{DElia:2016kpz}
M.~D'Elia, M.~Mariti, {Effect of Compactified Dimensions and Background
  Magnetic Fields on the Phase Structure of SU(N) Gauge Theories}, Phys. Rev.
  Lett. 118~(17) (2017) 172001.
\newblock \href {http://arxiv.org/abs/1612.07752} {\path{arXiv:1612.07752}},
  \href {https://doi.org/10.1103/PhysRevLett.118.172001}
  {\path{doi:10.1103/PhysRevLett.118.172001}}.

\bibitem{Kogut:1974ag}
J.~B. Kogut, L.~Susskind, {Hamiltonian Formulation of Wilson's Lattice Gauge
  Theories}, Phys. Rev. D 11 (1975) 395--408.
\newblock \href {https://doi.org/10.1103/PhysRevD.11.395}
  {\path{doi:10.1103/PhysRevD.11.395}}.

\bibitem{Wilson:1974sk}
K.~G. Wilson, {Confinement of Quarks}, Phys. Rev. D 10 (1974) 2445--2459.
\newblock \href {https://doi.org/10.1103/PhysRevD.10.2445}
  {\path{doi:10.1103/PhysRevD.10.2445}}.

\bibitem{Gattringer:2010zz}
C.~Gattringer, C.~B. Lang, {Quantum chromodynamics on the lattice}, Vol. 788,
  Springer, Berlin, 2010.
\newblock \href {https://doi.org/10.1007/978-3-642-01850-3}
  {\path{doi:10.1007/978-3-642-01850-3}}.

\bibitem{Damgaard:1988hh}
P.~H. Damgaard, U.~M. Heller, {The U(1) Higgs Model in an External
  Electromagnetic Field}, Nucl. Phys. B 309 (1988) 625--654.
\newblock \href {https://doi.org/10.1016/0550-3213(88)90333-1}
  {\path{doi:10.1016/0550-3213(88)90333-1}}.

\bibitem{Al-Hashimi:2008quu}
M.~H. Al-Hashimi, U.~J. Wiese, {Discrete Accidental Symmetry for a Particle in
  a Constant Magnetic Field on a Torus}, Annals Phys. 324 (2009) 343--360.
\newblock \href {http://arxiv.org/abs/0807.0630} {\path{arXiv:0807.0630}},
  \href {https://doi.org/10.1016/j.aop.2008.07.006}
  {\path{doi:10.1016/j.aop.2008.07.006}}.

\bibitem{Buividovich:2009bh}
P.~V. Buividovich, M.~N. Chernodub, E.~V. Luschevskaya, M.~I. Polikarpov,
  {Lattice QCD in strong magnetic fields}, eCONF C0906083 (2009) 25.
\newblock \href {http://arxiv.org/abs/0909.1808} {\path{arXiv:0909.1808}}.

\bibitem{Kluberg-Stern:1983lmr}
H.~Kluberg-Stern, A.~Morel, O.~Napoly, B.~Petersson, {Flavors of Lagrangian
  Susskind Fermions}, Nucl. Phys. B 220 (1983) 447--470.
\newblock \href {https://doi.org/10.1016/0550-3213(83)90501-1}
  {\path{doi:10.1016/0550-3213(83)90501-1}}.

\bibitem{Morel:1984di}
A.~Morel, J.~P. Rodrigues, {How to Extract {QCD} Baryons From a Lattice Theory
  With Staggered Fermions}, Nucl. Phys. B 247 (1984) 44--60.
\newblock \href {https://doi.org/10.1016/0550-3213(84)90371-7}
  {\path{doi:10.1016/0550-3213(84)90371-7}}.

\bibitem{Yang:2023vsw}
J.-C. Yang, X.-G. Huang, {QCD on Rotating Lattice with Staggered Fermions},
  2023.
\newblock \href {http://arxiv.org/abs/2307.05755} {\path{arXiv:2307.05755}}.

\bibitem{Velasco:2022gaw}
E.~G. Velasco, B.~B. Brandt, F.~Cuteri, G.~Endr\H{o}di, G.~Mark\'o, {Anomalous
  transport phenomena on the lattice}, PoS LATTICE2022 (2023) 173.
\newblock \href {http://arxiv.org/abs/2212.02148} {\path{arXiv:2212.02148}},
  \href {https://doi.org/10.22323/1.430.0173} {\path{doi:10.22323/1.430.0173}}.

\bibitem{Clark:2006wp}
M.~A. Clark, A.~D. Kennedy, {Accelerating Staggered Fermion Dynamics with the
  Rational Hybrid Monte Carlo (RHMC) Algorithm}, Phys. Rev. D 75 (2007) 011502.
\newblock \href {http://arxiv.org/abs/hep-lat/0610047}
  {\path{arXiv:hep-lat/0610047}}, \href
  {https://doi.org/10.1103/PhysRevD.75.011502}
  {\path{doi:10.1103/PhysRevD.75.011502}}.

\bibitem{Wolff:2003sm}
U.~Wolff, {Monte Carlo errors with less errors}, Comput. Phys. Commun. 156
  (2004) 143--153, [Erratum: Comput.Phys.Commun. 176, 383 (2007)].
\newblock \href {http://arxiv.org/abs/hep-lat/0306017}
  {\path{arXiv:hep-lat/0306017}}, \href
  {https://doi.org/10.1016/S0010-4655(03)00467-3}
  {\path{doi:10.1016/S0010-4655(03)00467-3}}.

\bibitem{Gottlieb:1988gr}
S.~A. Gottlieb, W.~Liu, R.~L. Renken, R.~L. Sugar, D.~Toussaint, {Hadron Masses
  with Two Quark Flavors}, Phys. Rev. D 38 (1988) 2245.
\newblock \href {https://doi.org/10.1103/PhysRevD.38.2245}
  {\path{doi:10.1103/PhysRevD.38.2245}}.

\bibitem{Cucchieri:1996jm}
A.~Cucchieri, T.~Mendes, {Study of critical slowing down in SU(2) Landau gauge
  fixing}, Nucl. Phys. B Proc. Suppl. 53 (1997) 811--814.
\newblock \href {http://arxiv.org/abs/hep-lat/9608051}
  {\path{arXiv:hep-lat/9608051}}, \href
  {https://doi.org/10.1016/S0920-5632(96)00789-X}
  {\path{doi:10.1016/S0920-5632(96)00789-X}}.

\bibitem{Paciello:1991bd}
M.~L. Paciello, C.~Parrinello, S.~Petrarca, B.~Taglienti, A.~Vladikas, {SU(3)
  lattice gauge fixing with overrelaxation and Gribov copies}, Phys. Lett. B
  276 (1992) 163--167, [Erratum: Phys.Lett.B 281, 417 (1992)].
\newblock \href {https://doi.org/10.1016/0370-2693(92)90557-K}
  {\path{doi:10.1016/0370-2693(92)90557-K}}.

\bibitem{Endrodi:2023wwf}
G.~Endrodi, G.~Marko, {QCD phase diagram and equation of state in background
  electric fields} (9 2023).
\newblock \href {http://arxiv.org/abs/2309.07058} {\path{arXiv:2309.07058}}.

\bibitem{Bellwied:2016gtm}
R.~Bellwied, S.~Bors\'anyi, Z.~Fodor, J.~G\"unther, S.~D. Katz, K.~K. Szab\'o,
  C.~Ratti, A.~Pasztor, {Fluctuations and correlations in finite temperature
  QCD}, PoS ICHEP2016 (2016) 369.
\newblock \href {https://doi.org/10.22323/1.282.0369}
  {\path{doi:10.22323/1.282.0369}}.

\end{thebibliography}
\bibliographystyle{elsarticle-num}

\end{document}